\documentclass[aps,pra,twocolumn,showpacs,groupedaddress]{revtex4}

\usepackage[dvips]{graphicx}

\begin{document}
\preprint{CAOUS-DW}

\title{Asymmetric double-well potential for single-atom interferometry}

\author{A. I. Sidorov}
\email[]{asidorov@swin.edu.au}
\author{B. J. Dalton}
\author{S. M. Whitlock}
\author{F. Scharnberg}
\altaffiliation[Also at ]{IQO, University of Hannover}

\affiliation{ARC Centre of Excellence for Quantum-Atom Optics and
Centre for Atom Optics and Ultrafast Spectroscopy, Swinburne
University of Technology, Melbourne, Victoria 3122, Australia}

\date{\today}

\begin{abstract}
We consider the evolution of a single-atom wavefunction in a
time-dependent double-well interferometer in the presence of a
spatially asymmetric potential. We examine a case where a single
trapping potential is split into an asymmetric double well and then
recombined again. The interferometer involves a measurement of the
first excited state population as a sensitive measure of the
asymmetric potential. Based on a two-mode approximation a Bloch
vector model provides a simple and satisfactory description of the
dynamical evolution. We discuss the roles of adiabaticity and
asymmetry in the double-well interferometer. The Bloch model allows
us to account for the effects of asymmetry on the excited state
population throughout the interferometric process and to choose the
appropriate splitting, holding and recombination periods in order to
maximize the output signal. We also compare the outcomes of the
Bloch vector model with the results of numerical simulations of the
multi-state time-dependent Schr\"{o}dinger equation.

\end{abstract}



\maketitle
\section{Introduction}
The evolution of a quantum system in a double-well potential has
been the subject of numerous theoretical studies. These include
treatments of Josephson-like oscillations \cite{Javanainen86,
Jack96}, dynamic splitting \cite{Menotti01} and interference
\cite{Castin97} of Bose-Einstein condensates (BECs). The
interpretation of these effects is based on the approach
\cite{Javanainen96}, in which interference patterns are seen to
evolve as a result of successive boson measurements which do not
identify the originating condensate. The production of cold atoms
and BEC in microtraps on atom chips \cite{Folman02, Haensel01a,
Ott01} and in micro-optical systems \cite{Dumke02} has stimulated a
great interest towards novel implementations of atom interferometers
\cite{Hinds01, Haensel01b, Andersson02, Stickney02, Negretti04,
Kreutzmann04} that are based on the use of double-well potentials.
Double-well atom interferometers (DWAI) of both the single-atom and
the BEC varieties are well suited to implementation on an atom chip.
Here microfabricated structures allow a precise control on a
sub-micron scale over the splitting and merging processes. The key
processes of splitting \cite{Cassettari00, Muller00, Esteve05} and
merging \cite{Haensel01c} of cold atomic clouds and even
interference of a BEC after splitting in a double well \cite{Shin04,
Schumm05, Shin05} have been already demonstrated. Although the
implementation of a DWAI using a BEC can lead to a $\sqrt{N}$-fold
enhancement in precision measurements \cite{Kasevich02}, phase
diffusion \cite{Javanainen97} associated with mean field effects is
of concern \cite{Shin04}. Double-well interferometry with a single
atom can allow a longer measurement time and in this regard has a
potential advantage. An on-chip single-atom interferometer can be
integrated with the source of atoms in a ground state - the
Bose-Einstein condensate - and be used for sensitive measurements of
gravitational fields. DWAI may also be applied to measure
collisional phase shift induced by the atom-atom interaction, which
is useful for quantum computation processes \cite{Calarco00}.

Two proposed schemes of a single-atom DWAI involve time-dependent
transverse \cite{Hinds01} and axial \cite{Haensel01b} splittings of
a trapping potential. An atom is initially prepared in the ground
state of a single symmetric trapping potential, which is then split
into a symmetric double well. A spatially-asymmetric potential is
then applied and non-adiabatic evolution leads to transitions
between ground and excited states. The asymmetry is then switched
off and the double well is recombined into the original potential.
The population of the excited state measures the effect of the
asymmetric potential. The DWAI can be considered as a quantum-state
Mach-Zehnder interferometer where the evolution of the quantum state
via the two separated wells is analogous to the propagation of an
optical field via two pathways.

However, these treatments ignore the effect of asymmetry during the
splitting and merging stages. In reality asymmetry is always present
and could be the result of imperfect horizontal splitting
(introducing a gravity-based asymmetry), external spatially-variable
magnetic and electric fields or different left and right trap
frequencies. We show that the presence of small asymmetries have
dramatic consequences on the interferometric process. We have
produced a simple model in terms of a Bloch vector evolution that
enables us to consider a splitting-holding-merging sequence
involving a double well and takes into account the presence of
asymmetry at all stages. The two key parameters are the energy gap
between the lowest two states of the symmetric component of the
trapping potential and an asymmetry parameter, which is related to
matrix elements of the asymmetric component of the potential.
Non-adiabatic evolution only occurs during the splitting and
recombining stages when the torque vector changes much more rapidly
compared to the Larmor precession of the Bloch vector. It is
important that the torque vector remains constant during the holding
stage, and that this period is long compared to the splitting and
recombination times. In this case the final excited state population
is a sinusoidal function of the holding time, with a period
determined via the asymmetry parameter.

In this paper we consider the dynamics of a single atom in an
asymmetric DWAI, with the basic theoretical framework being outlined
in Section II. Using a two-mode approximation we develop a Bloch
vector model for time-dependent DWAI (Section III) that provides an
adequate description of the dynamics of the splitting, holding and
recombination processes in the presence of an asymmetric component,
and is then used (Section IV) in describing the interferometric
process. The validity of the two-mode approach is explored in
Section V through the comparison of predictions of the Bloch vector
model with the results of direct numerical simulations of the
time-dependent multi-mode Schr{\"{o}}dinger equation. A discussion
of results follows in Section VI and includes a novel scheme to
measure the population of the excited state. Theoretical details are
dealt with in the Appendix.

\section{Theoretical Frame}
In general, the evolution of a single atom in an interferometer must
be described using a three dimensional quantum treatment. However,
for a system of cylindrical symmetry (as is present in typical atom
chip experiments) it is possible to ignore excitations associated
with the two tightly confined dimensions, as long as the dynamics
throughout the process is restricted to the dimension of weak
confinement (longitudinal splitting). In this system it is possible
to reduce the quantum treatment to that of a one dimensional
problem.

We consider the one-dimensional evolution of a single atom system
due to a time-dependent Hamiltonian $\widehat{H}(t)$ that can be
written as the sum of a symmetric Hamiltonian $\widehat{H_{0}}(t)$
and an asymmetric potential $\widehat{V_{as}}(\widehat{x})$
\begin{eqnarray}
\widehat{H}(t) &=&\widehat{H_{0}}(t)+\widehat{V_{as}}(\widehat{x})
\nonumber\\
&=&\frac{\widehat{p}^{2}}{2}+\widehat{V_{0}}(\widehat{x},t) +
\widehat{V_{as}}(\widehat{x}),  \label{Eq.Hamilt} \\
\widehat{V_{0}}(\widehat{x},t) &=&\left\{1+\left[\beta (t)-\frac{\widehat{x}^{2}}{2}%
\right]^{2}\right\}^{1/2},  \label{Eq.AsymmPotential}
\end{eqnarray}
where a specific form for the symmetric potential $\widehat{V_{0}}$
is chosen \cite{Pulido03}. The Hamiltonian and other physical
quantities have been written in dimensionless quantum harmonic
oscillator units associated with atomic mass $m$ and angular
frequency $\omega_{0}$. With the original quantities denoted by
primes we have
\begin{eqnarray}
\widehat{x} &=&\frac{\widehat{x}^{\prime}}{a_{0}},\quad
\widehat{p} =\frac{a_{0}}{\hbar }\,\widehat{p}^{\prime },  \label{Eq. norm} \\
t &=&\omega _{0}\,t^{\prime}, \quad   a_{0} =\sqrt{\frac{\hbar
}{m\,\omega_0}}.  \nonumber
\end{eqnarray}
The dimensionless Hamiltonians, potentials and energies are obtained
by dividing the original quantities by $\hbar \omega _{0}$.
$\widehat{V_{as}}$ will be taken as a linear function of
$\widehat{x}$.

The symmetric potential depends on a time-dependent splitting
parameter $\beta$, whose change from zero to a large value and back
to zero again conveniently describes the splitting, holding and
recombination processes with periods $T_{s}$, $T_{h}$ and $T_{r}$
respectively. For zero $\beta$ the symmetric potential involves a
single quartic well. When it is large a double harmonic well appears
with a separation between minima of $2\sqrt{2\beta}$. For zero
$\beta$ and for large $x$ the symmetric potential approximates that
for a quantum harmonic oscillator with frequency $\omega_{0}$ and
mass $m$. Key results in the paper would still apply if other
suitable forms for the symmetric potential are used.

We denote the eigenvectors of $\widehat{H}$ as $\vert
\phi_{i}\rangle$ and their energy eigenvalues as $E_{i}$, where
$i=0,1,2,..$ and $E_{i+1}>E_{i}$. The corresponding quantities for
the symmetric component of the Hamiltonian, $\widehat{H_{0}}$, will
be denoted $\vert S_{i}\rangle$ and $E_{S\, i}$. Both sets of
eigenvectors are orthonormal, and all energies and eigenvectors are
time dependent. $\widehat{H_{0}}$ is symmetric and the ground state
$\vert S_{0}\rangle$ is symmetric and denoted as $\vert S\rangle$,
whilst the first excited state $\vert S_{1}\rangle$ is antisymmetric
and denoted as $\vert AS\rangle$. Their energies are denoted $E_{S}$
and $E_{AS}$. The one-dimensional nature of the system allows real
eigenfunctions $\phi_{i}(x)$, $S_{i}(x)$ to be chosen. In this case
the geometric phase \cite{Berry84} is zero.

We can illustrate the general behavior of the lowest few energy
eigenvalues ($E_{0},\,E_{1},${\small \ }$E_{2},\,E_{3},..$) of the
total Hamiltonian $\widehat{H}$ as the splitting parameter is
increased from zero to a large value and back. At the beginning and
the end of the process when $\beta$=0 the energy eigenvalues are
well separated. Here the effect of asymmetry $\widehat{V_{as}}$ is
small and the eigenvalues and eigenvectors resemble those for the
symmetric Hamiltonian. When the splitting parameter increases and
the trapping potential changes to a double well, pairs of
eigenvalues ($E_{0}$ and $E_{1}$, $E_{2}$ and $E_{3}$, etc) become
very close. At this stage the quantum system is very sensitive to
the presence of $\widehat{V_{as}}$ which breaks the symmetry, allows
transitions between $\vert \phi_{0}\rangle$ and $\vert
\phi_{1}\rangle$ to occur and causes the eigenvectors $\vert
\phi_{0}\rangle$ and $\vert \phi_{1}\rangle$ (as well as $\vert
\phi_{2}\rangle$ and $\vert \phi_{3}\rangle$) to be localized in the
individual wells in this far split regime.

Initially, the atom is prepared in the lowest energy eigenstate
$\vert \phi_{0}\rangle$ of the single well. Transitions to excited
states are suppressed if the time scale for splitting and
recombination is much longer than the inverse frequency gap between
the relevant states. The energy gap between $E_{0}$ and $E_{2}$ is
always larger than the gap between $E_{0}$ and $E_{1}$, and by
choosing appropriate time scales we can isolate the two lowest
energy eigenstates ($\vert \phi_{0}\rangle$ and $\vert
\phi_{1}\rangle$) from higher excited states ($\vert
\phi_{2}\rangle$, $\vert \phi_{3}\rangle$, etc), but still allow for
transitions between the two lowest energy eigenstates to occur. As a
consequence the dynamics of the DWAI can be treated under the
two-mode approximation, in which only the two lowest energy
eigenstates of the total Hamiltonian $\widehat{H}$ and the symmetric
Hamiltonian $\widehat{H_{0}}$ need to be considered. In this case
the first excited state probability (a measurable quantity) can vary
from zero to one. A proposal for measuring the excited state
population is outlined in Section VI.

Using the two-mode approximation expressions for the lowest two
energy eigenvalues ($E_{0}$, $E_{1}$) and eigenvectors ($\vert
\phi_{0}\rangle$, $\vert \phi_{1}\rangle$) for the Hamiltonian
$\widehat{H}$ will be obtained. A standard matrix mechanics approach
will be used, but instead of using the symmetric potential energy
eigenvectors $\vert S\rangle$, $\vert AS\rangle$ as basis vectors,
we use the left, right (L-R) basis vectors $\vert L\rangle$, $\vert
R\rangle$, which are defined by
\begin{eqnarray}
\left\vert \,L\,\right\rangle  &=&\frac{1}{\sqrt{2}}(\left\vert
\,S\,\right\rangle +\left\vert \,AS\,\right\rangle ),  \label{Eq.LRStates} \\
\left\vert \,R\,\right\rangle  &=&\frac{1}{\sqrt{2}}(\left\vert
\,S\,\right\rangle -\left\vert \,AS\,\right\rangle ). \nonumber
\end{eqnarray}

The states $\vert L\rangle$, $\vert R\rangle$ are orthonormal and
for large $\beta$ correspond to an atom localized in the left or
right well, respectively. However, even for a single well the $L-R$
basis is still applicable. The matrix for the Hamiltonian
$\widehat{H}$ in the $L-R$ basis is given by
\begin{equation}
\lbrack \widehat{H}]^{L-R}=\frac{1}{2}(E_{S} + E_{AS})
\left(\begin{array}{cc}1 & 0 \\0 & 1\end{array}\right) +\frac{1}{2}
\left(\begin{array}{cc} -V_{as} & -\Delta_{0}
\\-\Delta_{0} & +V_{as}\end{array}\right) ,  \label{Eq.HamiltMatrix LR}
\end{equation}
where the order of the columns and rows is $L,R$ and we define the
convenient real quantities
\begin{eqnarray}
\Delta_{0} &=&E_{AS}- E_{S},  \label{Eq. Delta0} \\
V_{as} &=&\langle R|\,\widehat{V_{as}} \,|R\rangle - \langle
L|\,\widehat{V_{as}}\,|L\rangle  \nonumber \\
&=&-(\langle S|\,\widehat{V_{as}}\,|AS\rangle +\langle
AS|\,\widehat{V_{as}}\,|S\rangle ). \label{Eq. Vas 1}
\end{eqnarray}

The derivation of the Hamiltonian matrix uses the symmetry
properties of $\vert S\rangle$, $\vert AS\rangle$ and the reality of
the eigenfunctions. The total energy for the symmetric Hamiltonian
is given by $E_{S} + E_{AS}$, and the energy gap is given by
$\Delta_{0}$. The quantity $V_{as}$ describes the asymmetry of the
system, and would be zero if the Hamiltonian was symmetric. The
second equation relates $V_{as}$ to off-diagonal elements of the
asymmetric contribution to the Hamiltonian, indicating its role in
causing transitions between the eigenstates $\vert S\rangle$, $\vert
AS\rangle$ of the symmetric Hamiltonian. The Hamiltonian matrix
(\ref{Eq.HamiltMatrix LR}) is analogous to that for a two-level atom
interacting with a monochromatic light field \cite{Allen75}. The
symmetric Hamiltonian transition frequency $\Delta_{0}$ is analogous
to the Rabi frequency, whilst the quantity $V_{as}$ is analogous to
the detuning.

The energy eigenvalues for the total Hamiltonian $\widehat{H\text{
}}$ are obtained from the determinental equation as the eigenvalues
of the matrix $[\widehat{H\text{ }}]^{L-R}$, and are given by
\begin{eqnarray}
E_{0} &=&\frac{1}{2}\,(E_{\,S}+E_{\,AS})-\frac{1}{2}
\,\Delta,   \label{Eq.E01} \\
E_{1} &=&\frac{1}{2}\,(E_{\,S}+E_{\,AS})+\frac{1}{2} \,\Delta,
\nonumber
\end{eqnarray}
where the quantity $\Delta$ gives the energy gap for the total
Hamiltonian $\widehat{H}$ and is defined by
\begin{eqnarray}
\Delta  &=&\sqrt{\,\Delta_{0}^{2}+V_{as}^{\,2}} = E_{1}-E_{0}.
\label{Eq.Delta Gap}
\end{eqnarray}
In terms of the laser-driven two-level atom analogy, $\Delta$ would
be analogous to the generalized Rabi frequency.

The orthonormal energy eigenvectors for the total Hamiltonian
$\widehat{H}$ are given by
\begin{eqnarray}
\left\vert \,\phi _{0}\right\rangle
&=&\sqrt{\frac{1+V}{2}\,}\left\vert \,L\,\right\rangle
+\sqrt{\frac{1-V}{2}\,}\left\vert \,R\,\right\rangle,
\label{Eq.Eigenstates 01} \\
\left\vert \,\phi _{1}\right\rangle
&=&\sqrt{\frac{1-V}{2}\,}\left\vert \,L\,\right\rangle
-\sqrt{\frac{1+V}{2}\,}\left\vert \,R\,\right\rangle , \nonumber
\end{eqnarray}
where the effect of asymmetry is now represented by the variable
\begin{equation}
V=\frac{V_{as}}{\Delta}.  \label{Eq.V}
\end{equation}

On substituting for $\vert L\rangle ,$ $\vert R\rangle$ the
eigenvectors for the total Hamiltonian can be related to those for
the symmetric Hamiltonian $\widehat{H_{0}}$. At the beginning and
end of the interferometer process we find that the asymmetry
parameter $V_{as}$ is small compared to the symmetric energy gap
$\Delta_{0}$. For $V\ll 1$ the eigenvectors $\vert \phi_{0}\rangle$,
$\vert \phi_{1}\rangle$ become similar to $\vert S\rangle$ and
$\vert AS\rangle$ respectively. For $V_{as}\gg \Delta_{0}$
($V\simeq1$), the eigenvectors $\vert \phi_{0}\rangle$, $\vert
\phi_{1}\rangle$ become equal to $\vert L\rangle$, $\vert R\rangle$
respectively, the localized eigenvectors for the separate wells.

\begin{figure}[t]
\includegraphics[width=9cm]{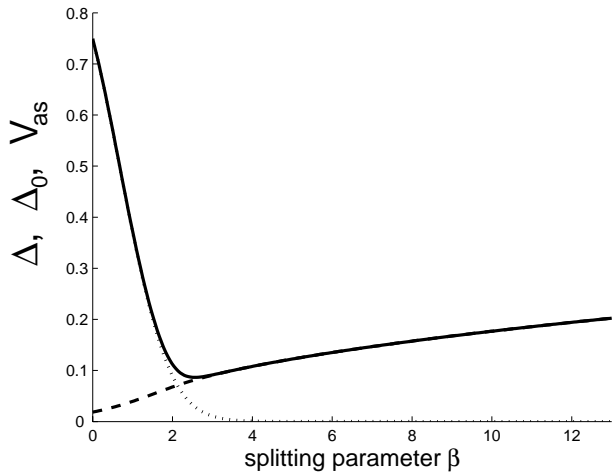}
\caption{Energy difference $\Delta$ between ground and first excited
states (solid line) for
$\widehat{V_{as}}\mathbf{=}0.02\,\widehat{x}$ as a function of the
splitting parameter $\beta$. Dotted line - energy difference
$\Delta_{0}$ for symmetric Hamiltonian, dashed line - asymmetry
quantity $V_{as}$.} \label{Fig1}
\end{figure}

\begin{figure}[t]
\includegraphics[width=6cm]{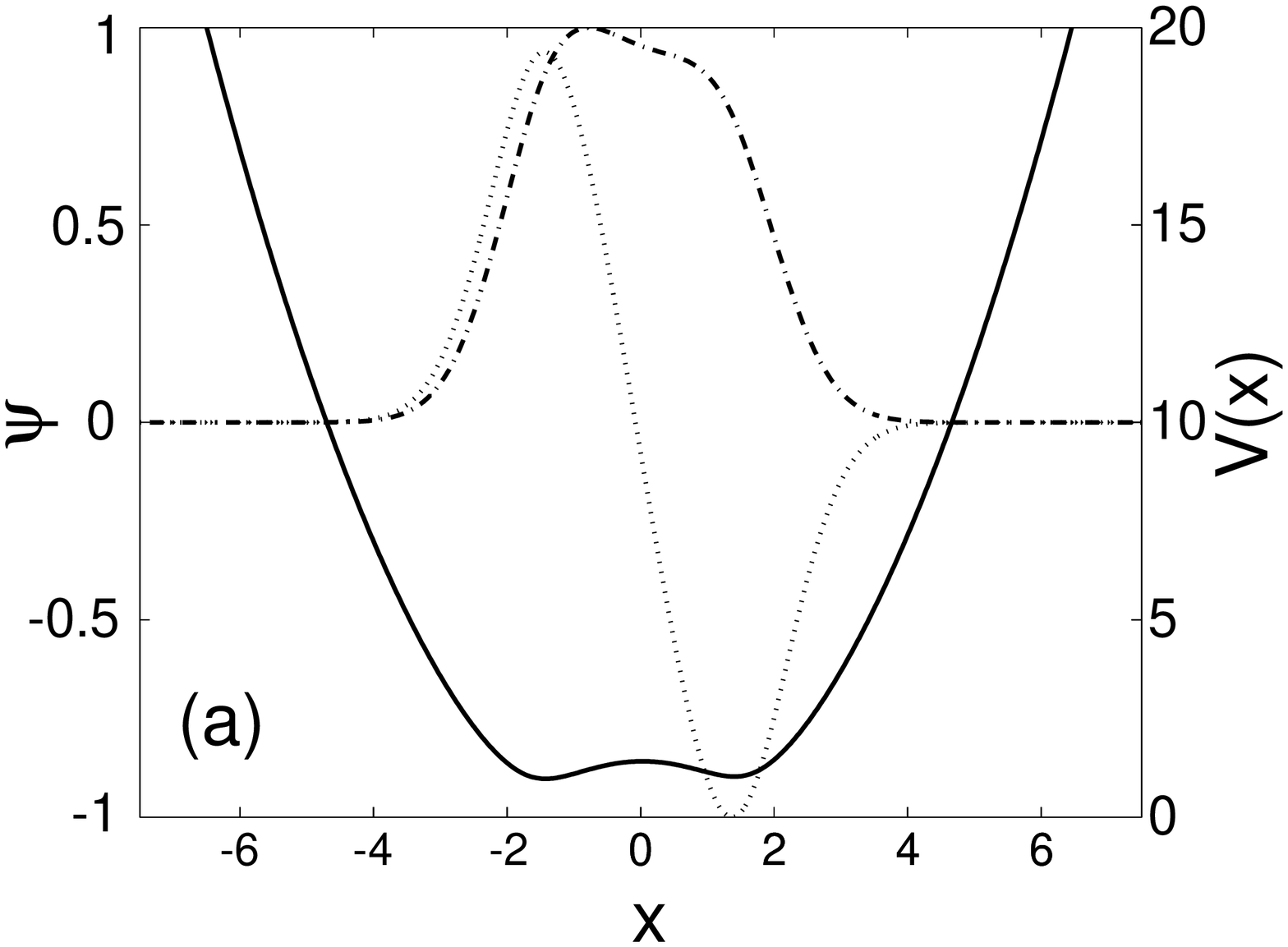}
\includegraphics[width=6cm]{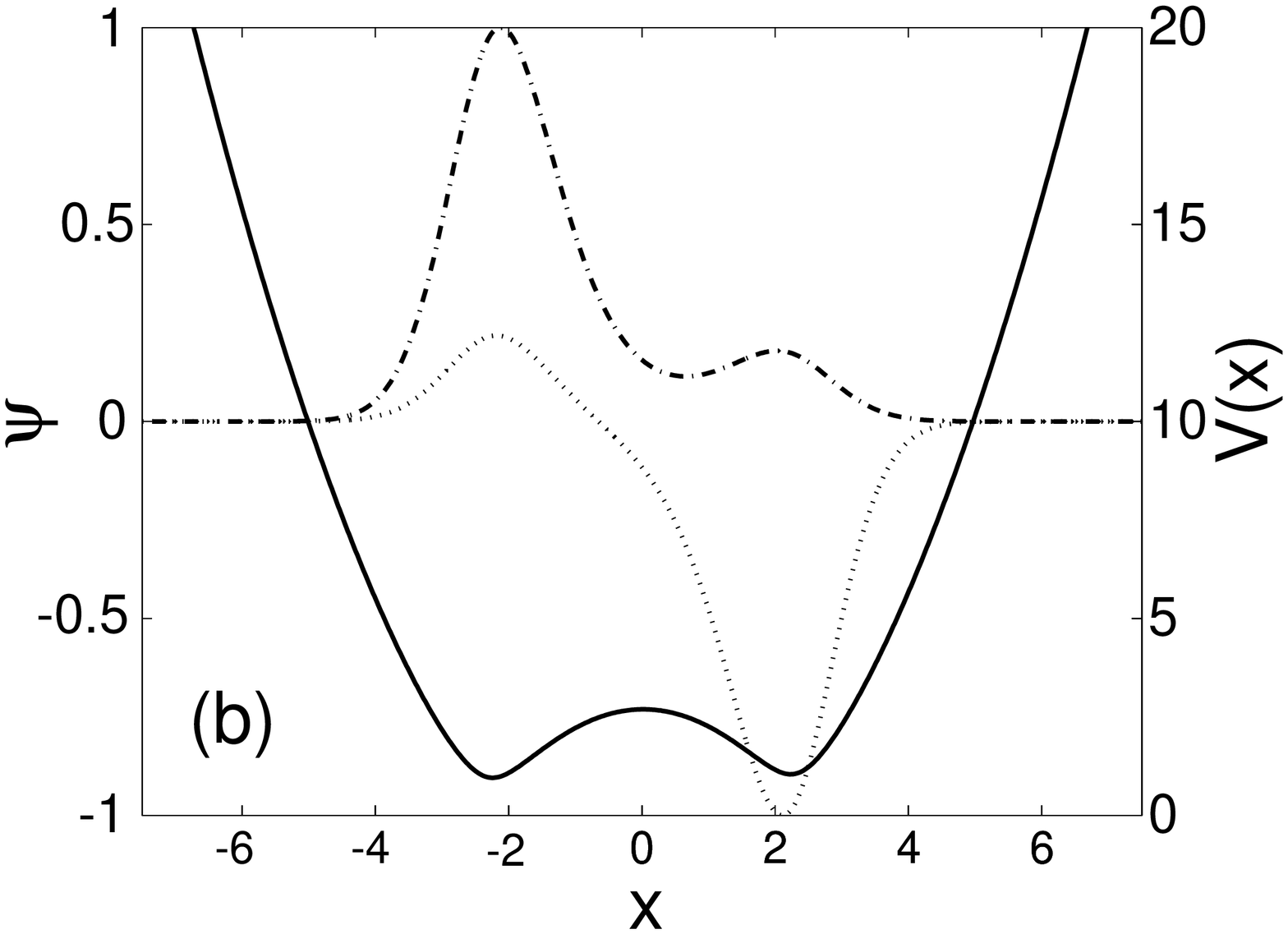}
\includegraphics[width=6cm]{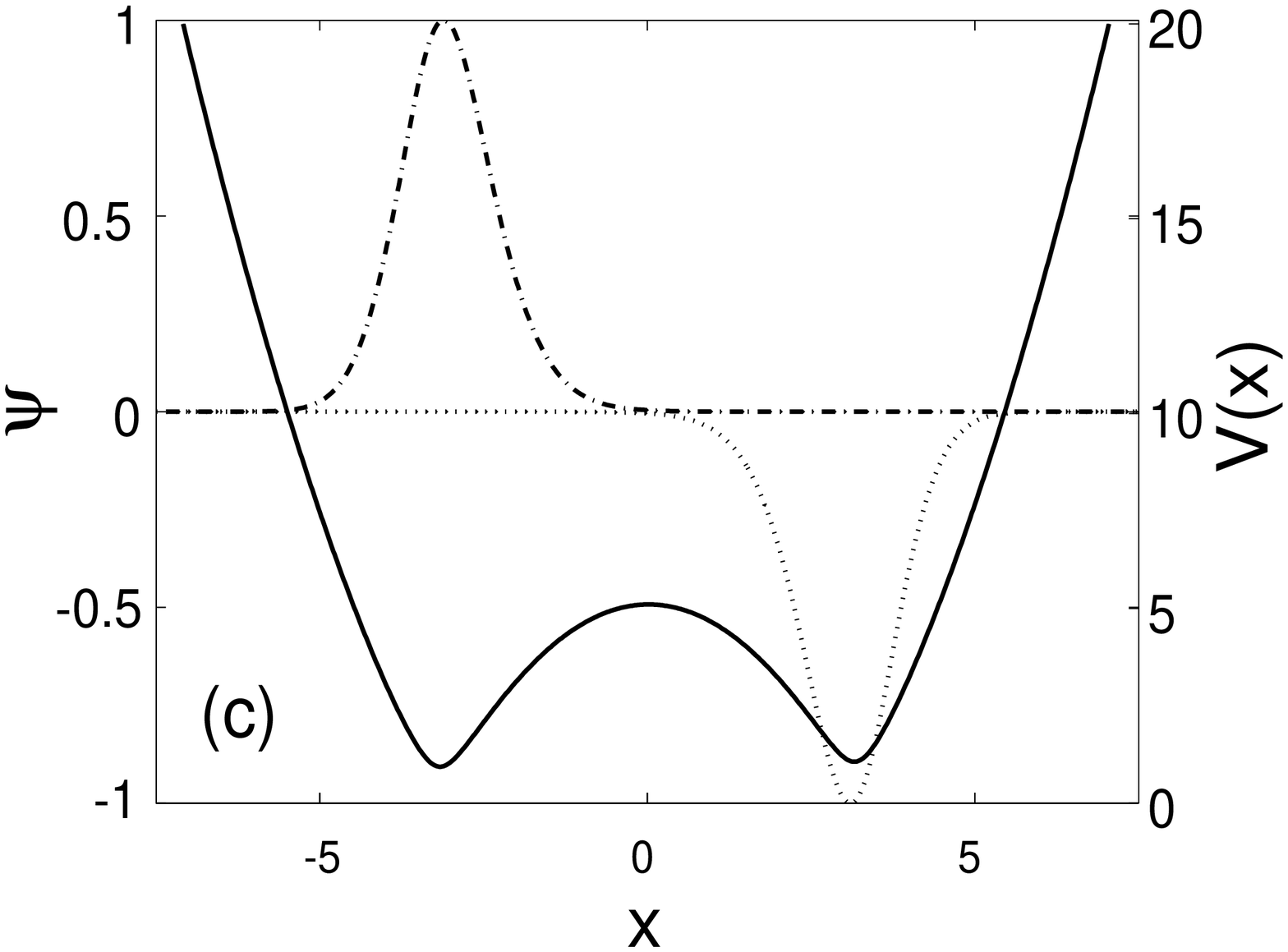}
\caption{Stationary eigenfunctions of the ground state
(dashed-dotted line) and the first excited state (dotted line) for
different values of the splitting parameter $\beta$ = 1 (a), 2.5 (b)
and 5 (c). The potential $V(x)$ is shown as the solid line.}
\label{Fig2}
\end{figure}

The behavior of the quantities $\Delta ,V_{as}$ and $\Delta_{0}$ as
the splitting parameter $\beta $ is changed is shown in
Fig.~\ref{Fig1} for the case where the asymmetric potential
$V_{as}(x)$ varies linearly with the coordinate $x$, specifically
with $\widehat{V_{as}}\mathbf{=}0.02\,\widehat{x}$. The symmetric
energy gap $\Delta_{0}$ becomes close to zero for $\beta \gtrsim 4$
and then the actual energy gap $\Delta$ is approximately given by
$V_{as}$. The energy eigenfunctions $\phi_{0}(x)$ and $\phi_{1}(x)$
for different splitting parameters $\beta$ are depicted in
Fig.~\ref{Fig2}, again with
$\widehat{V_{as}}\mathbf{=}0.02\,\widehat{x}$. The behavior of the
potential $V(x) = V_{0}(x)+V_{as}(x)$ is also shown. For small
$\beta$ [Fig.~\ref{Fig2}(a)] the potential is a single well and the
eigenfunctions are approximately symmetric and antisymmetric. For
larger $\beta$ [Fig.~\ref{Fig2}(c)] the potential is a double well,
which still appears to be symmetric. However, even with a small
asymmetry in the potential the eigenfunctions are no longer
symmetric and antisymmetric, but instead are each localized in
separate wells. This sensitivity of the eigenfunctions to a small
asymmetry is critical to the performance of the present
interferometer.

\section{Bloch Vector Model}
We can express a general time-dependent normalized state vector
$\vert \Psi\rangle$ as a quantum superposition of the states $\vert
L\rangle$ and $\vert R\rangle$
\begin{equation}
\vert \,\Psi (t)\,\rangle =C_{L}\,\vert \,L\,\rangle +C_{R}\,\vert
\,R\,\rangle . \label{Eq.State Vector Expn}
\end{equation}

Our interferometer will be described in terms of the Bloch vector
and its dynamics determined via Bloch equations. We now introduce
Pauli spin operators and the Bloch vector. Time-dependent Pauli spin
operators $\widehat{\sigma_{a}}$ $(a=x,y,z)$ are defined in the
Schr\"{o}dinger picture
\begin{eqnarray}
\widehat{\sigma_x} &=&(|R\rangle \,\langle
L|\,+\,|L\rangle \,\langle R|),  \nonumber \\
\widehat{\sigma_y} &=&\frac{1}{i}\,(|R\rangle \,\langle
L|\,-\,|L\rangle \,\langle R|),  \label{Eq. Sxyz} \\
\widehat{\sigma_z} &=&(|R\rangle \,\langle R|\,-\,|L\rangle
\,\langle L|).  \nonumber
\end{eqnarray}

From its matrix representation in the $L-R$ basis
(\ref{Eq.HamiltMatrix LR}), the dimensionless Hamiltonian operator
$\widehat{H}$ in the Schr\"{o}dinger picture can be expressed in
terms of the Pauli spin operators as
\begin{equation}
\widehat{H}=\frac{1}{2}\,(\Omega_{0}\,\widehat{1}+\Omega_{x}\,
\widehat{\sigma_{x}}+\Omega_{y}\,\widehat{\sigma_{y}} + \Omega
_{z}\,\widehat{\sigma_{z}}), \label{Eq. Ham Opr Pauli}
\end{equation}
where
\begin{eqnarray}
\Omega_0 &=&(E_S+ E_{AS}), \label{Eq. Omega} \\
\Omega_x &=&-\,\Delta_0, \quad \Omega_y = 0, \quad \Omega_z =
V_{as}. \nonumber
\end{eqnarray}
It is convenient to introduce a so-called torque vector, defined as
$\vec{\Omega} =(\Omega_x,\,\Omega_y,\,\Omega_z)$.

The Bloch vector is defined to have components which are the
expectation values of the Pauli spin operators
$\widehat{\sigma_{a}}$ in the quantum state $\vert \Psi\rangle$.
These components will be denoted as $\sigma_{a}$. Hence in the
Schr\"{o}dinger picture
\begin{equation}
\sigma_{a}=\left\langle\,\Psi(t)\,|\,\widehat{\sigma _{a}}(t)\,|\,
\Psi (t)\right\rangle \qquad (a=x,y,z).  \label{Eq. Bloch Vector 2}
\end{equation}
The Bloch vector is defined as $\vec{\sigma} =(\sigma_x,
\,\sigma_y,\,\sigma_z)$. The Bloch components are bilinear functions
of the amplitudes $C_{L}$ and $C_{R}$.

The evolution of the DWAI system is now described by a set of real
variables $\sigma_{x},\,\sigma_{y},\,\sigma_{z}$ and each of these
variables has a certain physical meaning. The component $\sigma_{z}$
is a measure of the imbalance of the atomic population of the
localized states $\vert L\rangle ,$ $\vert R\rangle$. The component
$\sigma_{x}$ is a measure of the atomic population imbalance between
the delocalized states $\vert S\rangle$, $\vert AS\rangle$, as can
be seen if the quantum state is expanded in the symmetric basis. For
$\sigma_{x}=+1$ all the population is in the symmetric state $\vert
S\rangle$, for $\sigma_{x}=-1$ it is all in the antisymmetric state
$\vert AS\rangle$. It is thus a measure of the excitation of the
first excited state in the unsplit trap regime.

Equations for the components of the Bloch vector can be obtained
from Heisenberg equations for the Pauli spin operators. The
derivation must take into account the present situation where the
Pauli spin operators are explicitly time dependent since the basis
vectors $\vert L\rangle ,$ $\vert R\rangle$ change with time. This
differs from the standard situation of time independent basis
vectors \cite{Allen75, Tannoudji77}. However, the additional term in
the Heisenberg equations can be shown to contribute zero to the
Bloch equations due to the two eigenfunctions in the symmetric basis
being real and having opposite symmetry (see Appendix). The Bloch
equations are given by ($d_{t} \equiv d/dt$)
\begin{eqnarray}
d_{t}\,\sigma_x &=&-\,V_{as}\,\sigma_y,  \nonumber \\
d_{t}\,\sigma_y &=&\,V_{as}\,\sigma_x+\,\Delta_0\,\sigma_z,  \label{Eq. Bloch Eqn}\\
d_{t}\,\sigma_z &=&-\,\Delta_0\,\sigma_y,  \nonumber
\end{eqnarray}
and can be solved numerically using the Runge-Kutta algorithm. In
vector notation the Bloch equations are
\begin{equation}
d_{t}\,\vec{\sigma} =\vec{\Omega}\times \vec{\sigma}. \label{Eq.
Bloch Eqns Torque}
\end{equation}

This form of the Bloch equations is a direct consequence of the
equivalence of the two-mode double-well interferometer to a spin
$1/2$ system. The Bloch vector precesses at the Larmor frequency
around the torque vector, which in detail is
\begin{equation}
\vec{\Omega}=(-\Delta_0,\,0,\,V_{as}). \label{Eq. Torque Vector}
\end{equation}

\begin{figure}
\includegraphics[width=4cm]{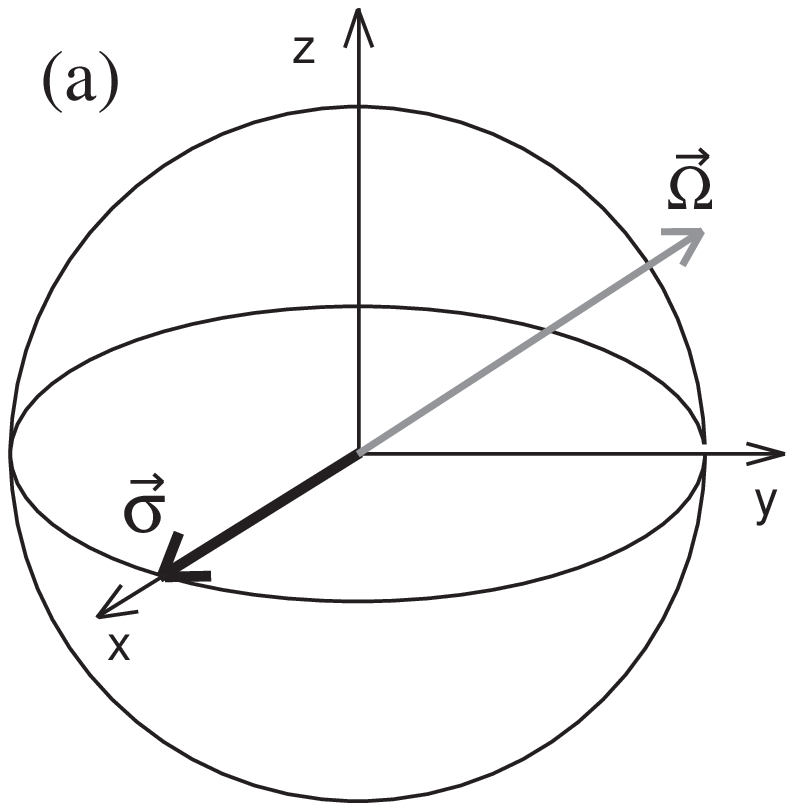}
\includegraphics[width=4cm]{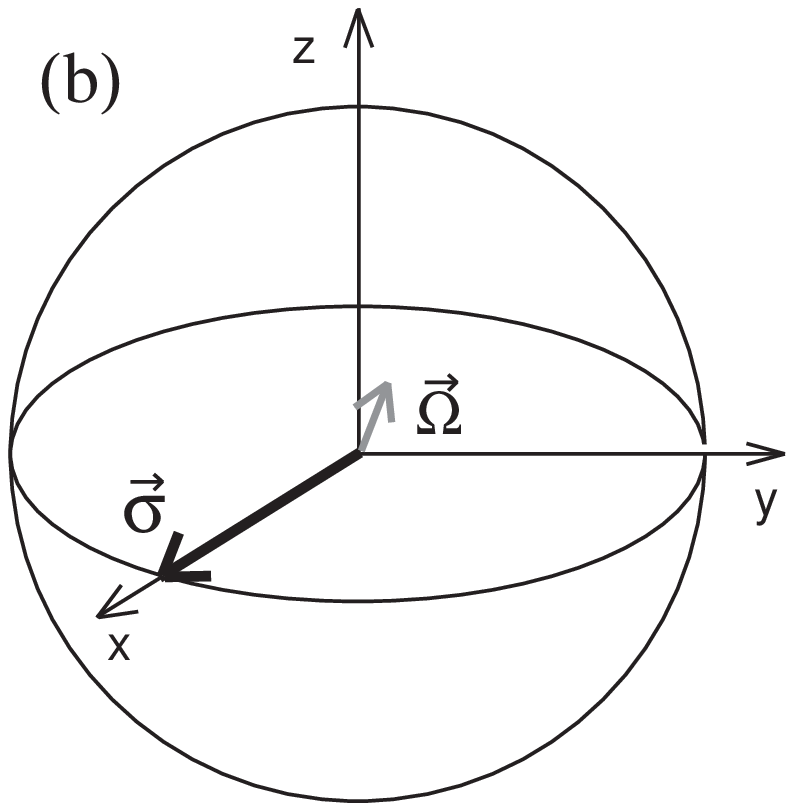}
\includegraphics[width=4cm]{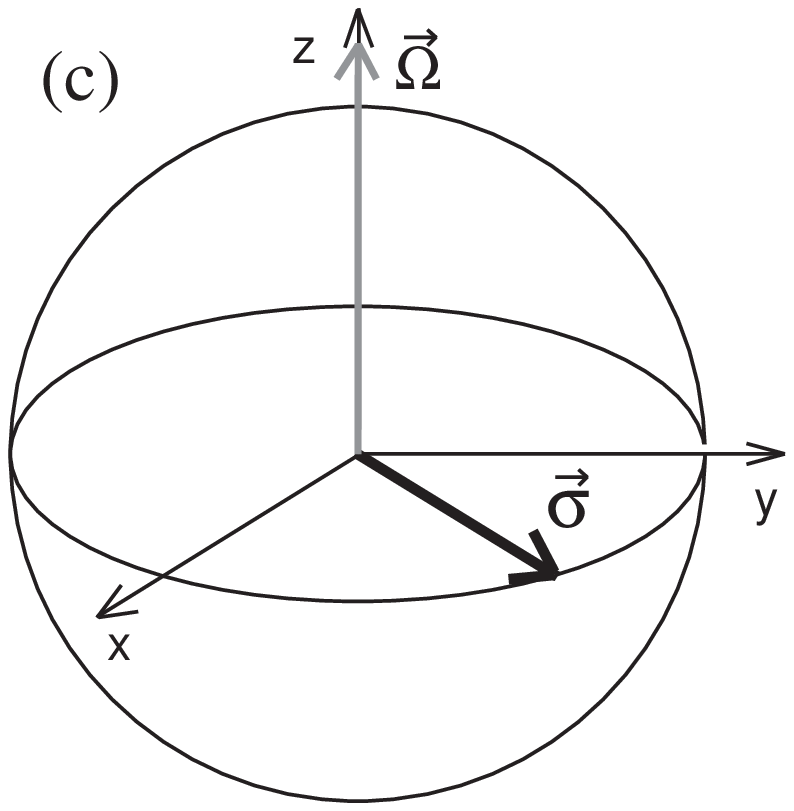}
\caption{Evolution of the Bloch vector $\vec{\sigma}$ and the torque
vector $\vec{\Omega}$ at different moments of the splitting stage:
(a) - at the beginning when $\Delta_{0} \gg V_{as}$ and
$\vec{\Omega} \approx (-\Delta_{0},\,0,\,0)$ (b) - when $\Delta_{0}
= V_{as}$, (c) - when $V_{as} \gg \Delta_{0}$ and $\vec{\Omega}
\approx (0,\,0,\,V_{as})$.} \label{Fig3}
\end{figure}

If there is no asymmetry, the $x$ component of the Bloch vector
remains unchanged, whilst its component in the $y-z$ plane just
rotates about the $x$ axis (Fig.~\ref{Fig3}).

\section{Model of a single-atom double-well atom interferometer}
In the single-atom interferometer under consideration the atom is
always located in a trapping potential, which changes from a single
well to a double well - which in general is slightly asymmetric -
and back again to the original single well. The interferometer is
used to measure the effects of this asymmetry, the cause of which
may be of measurable interest (e.g., as in a gravity gradiometer).
The atom is initially in the ground state $\vert \phi_{0}(0)\rangle$
of the original unsplit potential, and as $V_{as}$ is then small
compared to $\Delta_{0}, \vert \phi_{0}(0)\rangle$ is then
approximately the same as $| S(0)\rangle$. The population of the
excited state at the end of the recombination process is the
measurable interferometer output. The probability $P_{1}$ of finding
the atom in the upper energy eigenstate at any time is given by
\begin{equation}
P_{1}=|\langle \,\phi _1|\,\Psi \,\rangle |^{2}, \label{Eq.State 1
Probability}
\end{equation}
and this will remain zero unless an asymmetry is present together
with suitably short splitting and recombining stages for the
interferometer process - so that transitions occur between $\vert
\phi_{0}\rangle $ and $\vert \phi_{1}\rangle$ due to the presence of
$\widehat{V_{as}}$.

We find that
\begin{eqnarray}
P_1 &=&\frac{1}{2}(1+\sigma_z\,V-\sigma_x\,\sqrt{
1-V^{2}})  \nonumber \\
&=&\frac{1}{2}+\frac{1}{2{\Delta }}\,\vec{\Omega} \cdot
\vec{\sigma}. \label{Eq. State 1 Probability Torque}
\end{eqnarray}

At the beginning and end of the interferometer process $V\ll 1$ and
hence the probability $P_{1}$ only depends on the $x$ component of
the Bloch vector. The probability $P_{1}(T)$ thus depends on how
this component has changed from its initial value of 1. We can
therefore describe the dynamical behavior of the single atom
interferometer in terms of the evolution of the Bloch vector during
the splitting, holding and recombining stages. At the start of the
process the Bloch and torque vectors are antiparallel
[Fig.~\ref{Fig3}(a)] and approximately aligned with the $x$ axis.
For small values of the splitting parameter $\beta$ the absolute
value of the torque vector is mainly determined by the symmetric
energy gap $\Delta_{0}$ (Fig.~\ref{Fig1}) and its direction remains
along the -$x$ direction. The position of the Bloch vector remains
mostly along the $+x$ direction [Fig.~\ref{Fig3}(b)] during early
stages of the splitting process. When the splitting parameter is
increased further the decreasing energy gap $\Delta_{0}$ becomes
comparable with and later much smaller than the asymmetry quantity
$V_{as}$. As a result the torque vector rotates in a $x$-$z$ plane
until it is aligned along the $z$ direction [$\vec{\Omega} \approx
(0,\,0,\,V_{as})$]. It is important to make this change
non-adiabatically so that the Bloch vector does not follow the
torque rotation. If the Bloch vector were to follow the changes of
the torque vector adiabatically the atom will always stay in the
ground state and no interference would be observed.

During the holding stage the torque vector is constant and the Bloch
vector precesses around the torque vector with a constant angular
velocity $V_{as}$, and hence both the $x$ and $y$ components
oscillate with a period $2\pi/V_{as}$ [Fig.~\ref{Fig3}(c)]. In an
ideal double-well interferometer the splitting and recombination
stages are short and the value of the $x$ component does not change
much during these stages, so that $\sigma_x(T)$ (which defines the
final excited state population) is basically given by its value at
the end of the holding period. The simple behavior during the
holding period indicates that the excited state population would
have a period $2\pi/V_{as}$ considered as a function of holding
time. A similar description in terms of the evolution of a Bloch
vector also applies to the scheme described in Ref. \cite
{Haensel01b}, though the dynamical behavior of the Bloch vector is
different.

The behavior of the interferometer may also be described in terms of
time-dependent states $\vert L\rangle$, $\vert R\rangle$, which
during the holding period represent atoms localized in the left and
right wells. The interferometer process involves the transition
$\vert S(0)\rangle$ $\longrightarrow \vert AS(T)\rangle$, which
involves two pathways $\vert S(0)\rangle \longrightarrow \vert
L(T/2)\rangle \longrightarrow \vert AS(T)\rangle $ and $\vert
S(0)\rangle \longrightarrow \vert R(T/2)\rangle \longrightarrow
\vert AS(T)\rangle $, involving two possible localized intermediate
states associated with the left or right wells. The overall
transition amplitude is the sum of amplitudes for the two pathways,
and depending on the relative phase between these amplitudes either
constructive or destructive interference may occur. For maximum
contrast it is desirable that the magnitudes of the two partial
amplitudes be equal, so that during the holding period the
populations of the left and right well states should be about the
same. After optimal splitting the $z$ component of the Bloch vector
$\sigma _{z}$ should be kept close to zero during the holding
period, however a phase difference between the localized states
accumulates. Only at the end of the recombination stage this phase
is translated into the population of the excited state.

\section{Results of numerical simulations}
\begin{figure}[t]
\includegraphics[width=8cm]{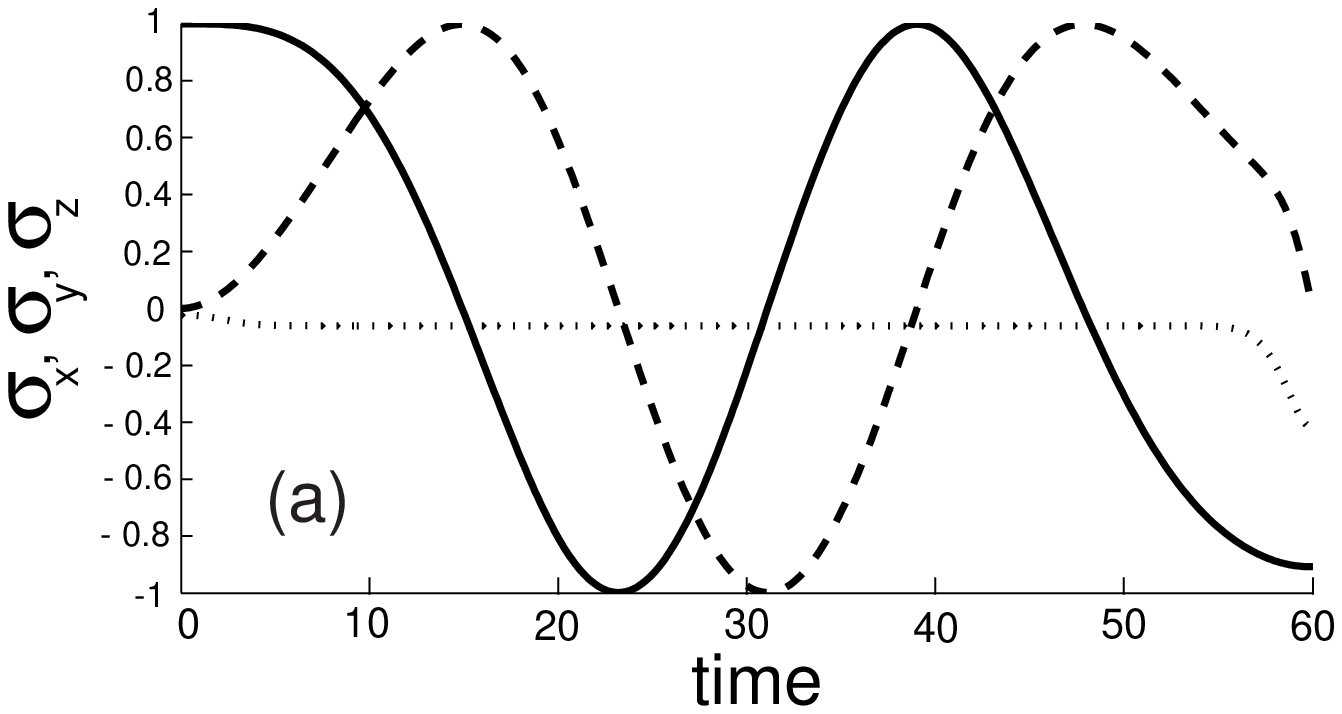}
\includegraphics[width=8cm]{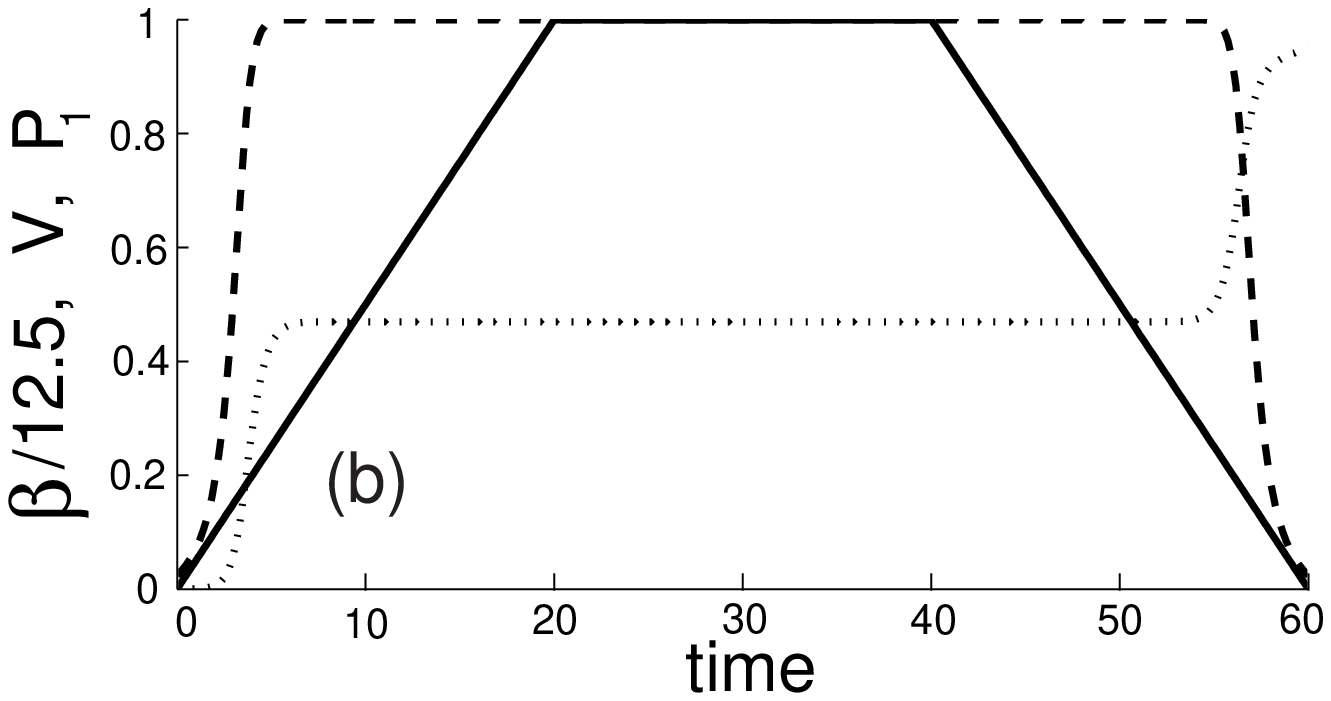}
\caption{(a) Time evolution of the Bloch vector components
$\sigma_{x}$ (solid line), $\sigma_{y}$ (dashed line) and
$\sigma_{z}$ (dotted line) for $T_{s} = T_{h} = T_{r}=20$ and
$\beta_{\max} = 12.5$; (b) Time evolution of the first excited state
population $P_{1}$ (dotted line), the asymmetry parameter $V$
(dashed line) and the splitting parameter $\beta$/$\beta_{max}$
(solid line).} \label{Fig4}
\end{figure}

We studied the evolution of a Bloch vector and the population of the
excited state by solving Equations (\ref{Eq. Bloch Eqn})
numerically. The splitting parameter $\beta $ is changed linearly
from zero up to a maximum $\beta _{\max }$ during the splitting
period. It is then held constant at $\beta _{\max }$ during the
holding period, and then changed linearly to zero during the
recombination period. The dynamical behavior of the Bloch vector
components is shown in Fig.~\ref{Fig4}(a) along with the time
dependence of the asymmetry parameter $V=V_{as}/ \Delta$, the
splitting parameter $\beta$ and the excited state population $P_{1}$
[Fig.~\ref{Fig4}(b)]. The parameters used are $\widehat{V_{as}} =
0.02\, \widehat{x}$ , $\beta_{\max } = 12.5$ and $T_{s}=20$,
$T_{h}=20$, $T_{r}=20$ in dimensionless harmonic oscillator units.
Here we observe complex oscillatory behavior for the $x$ and $y$
components of the Bloch vector which occurs during the splitting and
merging stages. During the holding stage they exhibit simple
periodic variations with frequency $V_{as}$ = 0.2. At the same time
the $z$ component develops a small negative value during splitting
and increases the absolute value even further during merging. The
$x$ component reaches a negative value of -0.9 at the end of the
process. This corresponds to an excited state population of 0.95 and
represents a case of constructive interference.

By monitoring the behavior of $P_1$ during the interferometric
process we can see when non-adiabatic evolution occurs. The
population $P_{1}$ changes from 0 to 0.47 [Fig.~\ref{Fig4}(b)] at
the beginning of the splitting process and does not reach the
optimal value 1/2 as a result of the non-zero $z$ component of the
Bloch vector. The variable $P_{1}$ does not change during the
adiabatic precession of the Bloch vector around the torque vector
during the rest of the splitting, holding and the beginning of
recombining stages. It again exhibits drastic changes in a short
period during the re-merging when the torque vector rotates rapidly
and the Larmor frequency is relatively small.

\begin{figure}[t]
\includegraphics[width=7cm]{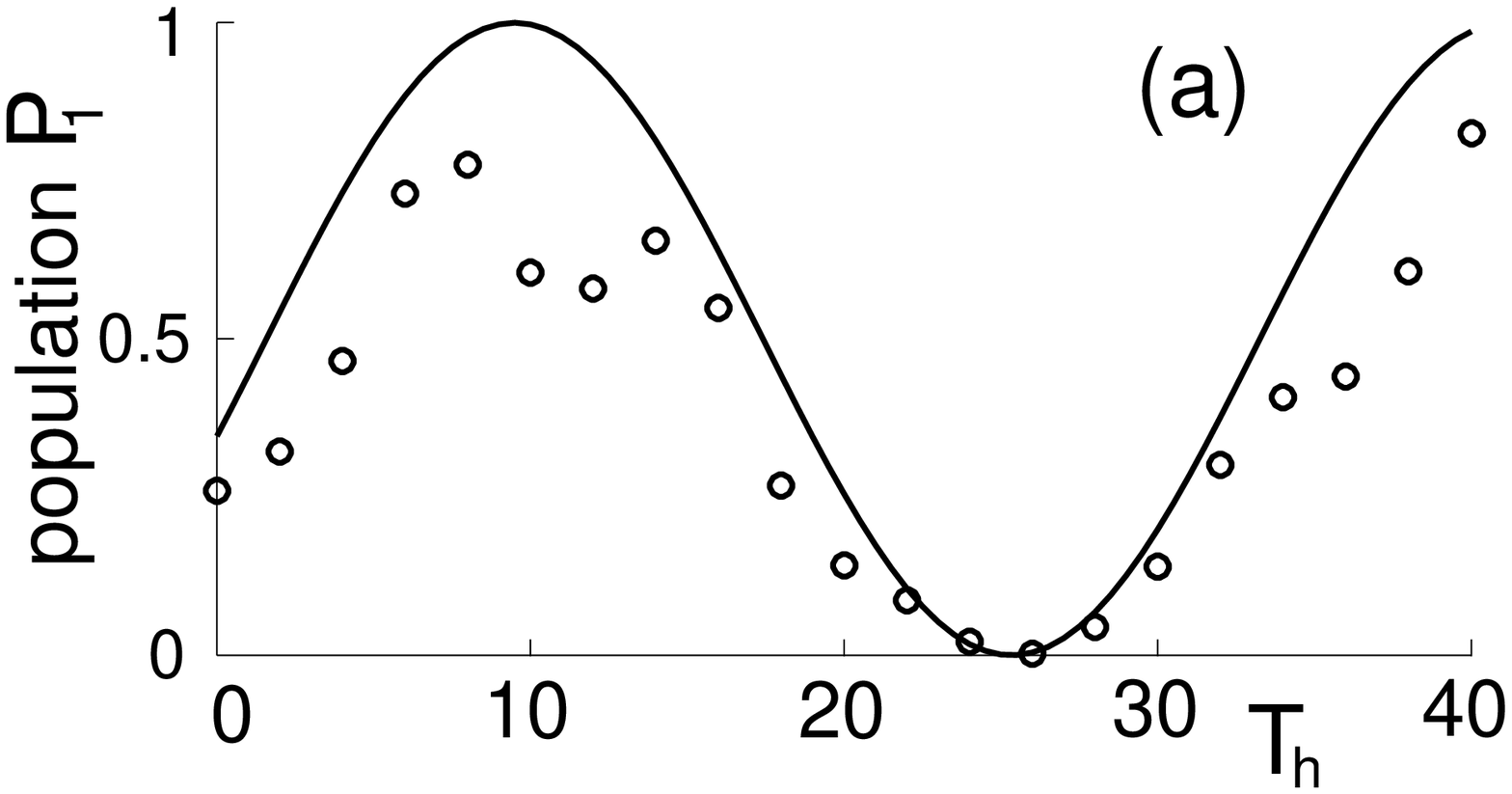}
\includegraphics[width=7cm]{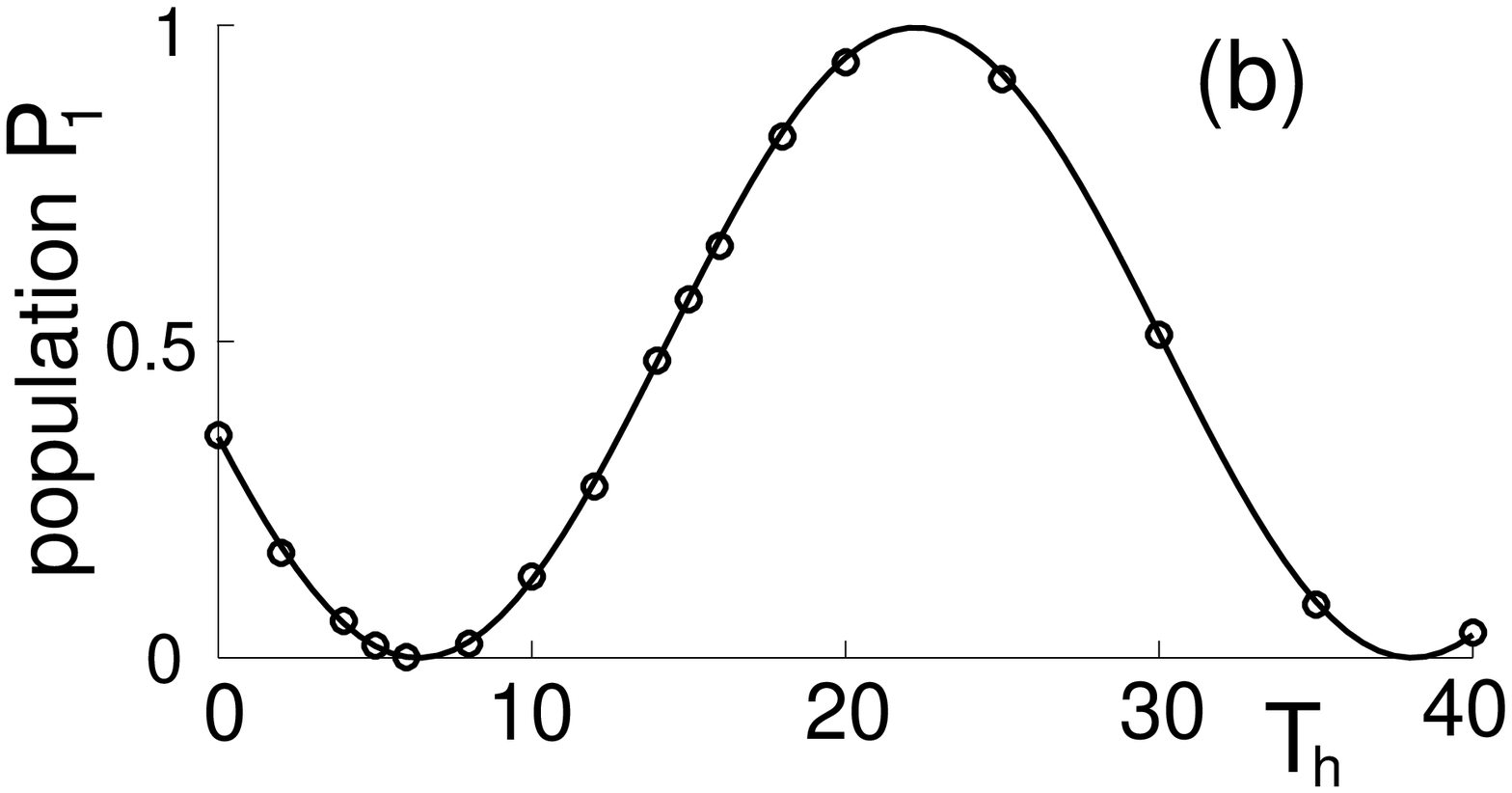}
\includegraphics[width=7cm]{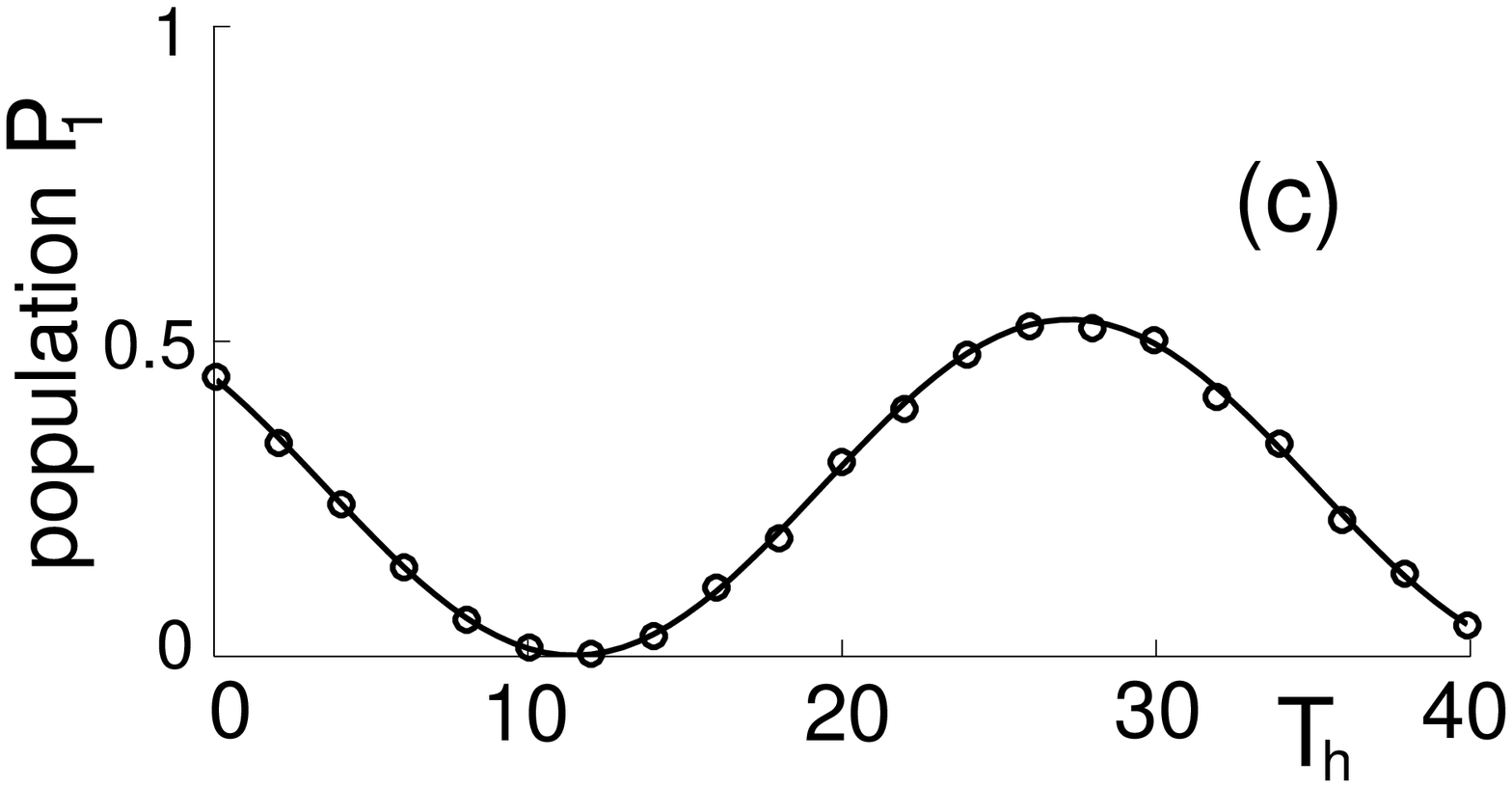}
\caption{Dependence of the first excited state population $P_{1}$ at
the end of the interferometric process on the duration of the
holding stage $T_{h}$ for various durations of the splitting and
recombining stages $T_{s}\,$ = $T_{r}$= 5 (a), 20 (b) and 200 (c).
Results of the Bloch vector model are represented by solid lines and
outcomes of full numerical simulations are presented by circles.}
\label{Fig5}
\end{figure}

It is tempting to limit the evolution of the Bloch vector and the
relevant phase accumulation during the splitting and merging stages
by making these stages shorter. However this can lead to excitations
of higher excited states. We have compared outcomes of the Bloch
vector model with the results of the numerical solution of a
multi-state Schr\"{o}dinger equation using the XMDS code \cite
{XMDS}. The behavior of the excited state population $P_{1}(T)$ at
the end of the interferometer process as a function of the holding
period $T_{h}$ is shown in Fig.~\ref{Fig5} for the parameters
$\widehat{V_{as}}=0.02\,\widehat{x}$ , $\mathbf{\beta }_{\max
}=12.5$, $T_{s}=T_{r}=5$ [Fig.~\ref{Fig5}(a)], $T_{s}=T_{r}=20$
[Fig.~\ref{Fig5}(b)] and $T_{s}=T_{r}=200$ [Fig.~\ref{Fig5}(c)]. In
all cases the sinusoidal behavior of the excited state population as
a function of the holding period can be seen. Situations ranging
from complete destructive interference to perfect constructive
interference are both present. For short splitting times
[Fig.~\ref{Fig5}(a)] we observed a discrepancy between the two
models. Multi-state numerical simulations indicate the presence of
populated higher energy states which the Bloch vector model ignores.
The full numerical calculations show an irregular high frequency
modulation of the fundamental frequency signal and a reduced maximum
population of the first excited state.

For splitting and merging times $T_{s}=T_{r}=20$
[Fig.~\ref{Fig5}(b)] both models show excellent agreement indicating
a simple sinusoidal variation of the first excited state population
with holding time. This simple behavior is also observed for long
splitting time [Fig.~\ref{Fig5}(c)] but with significantly reduced
amplitude of the oscillations. The reduced fringing is attributed to
the onset of adiabatic following of the Bloch vector during
splitting and recombination which is shown by the presence of a
non-zero $\sigma _{z}$ component [Fig.~\ref{Fig4}(a)]. We noticed
that our numerical solution of the Bloch equation is robust with
regard to the variations of the signal but is fragile regarding the
phase. The error was accumulated during the splitting stage as a
result of $V_{as}\neq{0}$ in a merged trap and will scale with the
splitting time.

\begin{figure}[t]
\includegraphics[width=8cm]{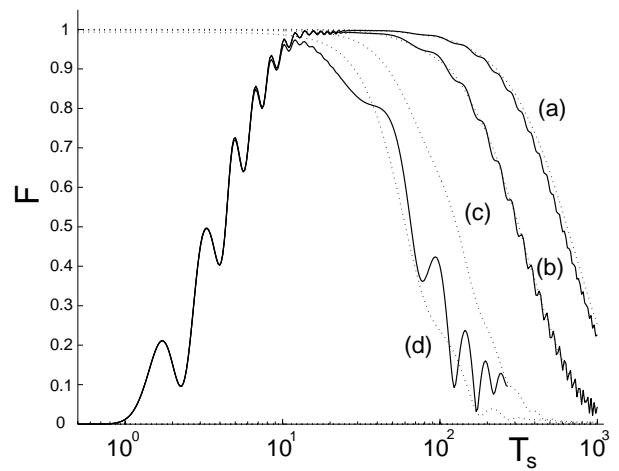}
\caption{Dependence of the filling factor $F$ on the duration of the
splitting stage $T_{s}$ for $\beta_{\max} = 12.5$ and various values
of asymmetry $\widehat{V_{as}} = 0.01\, \widehat{x}\, (a), 0.02\,
\widehat{x}\, (b), 0.05\, \widehat{x}\, (c)$ and $0.1\,
\widehat{x}\, (d)$.} \label{Fig6}
\end{figure}

In the asymmetric double-well potential the ground state
eigenfunction will predominantly occupy the lower well, and the
excited state eigenfunction will be localized in the upper well
[Fig.~\ref{Fig2}(c)]. In the slow splitting regime the onset of the
adiabatic evolution will lead to the unbalanced distribution of the
atomic wavefunction between the wells, which in turn leads to a
reduction in the measured signal. In application to interferometry
it can be seen as intrinsic which-way information when the atom will
predominantly follow one path after the splitting. In general
\begin{equation}
\left\vert \,\Psi\,\right\rangle =a\,\left\vert
\,\phi_{0}\,\right\rangle + b\,\left\vert \,\phi_{1}\,\right\rangle
+ \,\left\vert \,\phi_{i}\,\right\rangle , \label{Eq.combination}
\end{equation}
where $\left\vert \,\phi_{i}\,\right\rangle$ is a linear combination
of all other excited states. We define a filling factor
\begin{equation}
F = 2ab, \label{Eq.filling}
\end{equation}
which will describe the balance of ground and excited states
populations. The dependence of the filling factor on splitting time
is shown in Fig.~\ref{Fig6} for a splitting of $\beta = 12.5$ and
different asymmetries. The results of the Bloch model (dotted line)
and the multi-state numerical simulations (solid line) show good
agreement for splitting times $T_{s} \geq 20$. For shorter splitting
stages the two mode approximation fails and excitations into higher
modes take place. In the case of the high asymmetry
$\widehat{V_{as}} = 0.1\, \widehat{x}$ [Fig.~\ref{Fig6}(d)] we
observe a significant deviation between outcomes of two models. For
large values of the asymmetry frequency $V_{as}$ it is impossible to
adiabatically isolate two lower states and the transitions to higher
states have to be taken into account.

\section{Discussion and conclusions}
We have applied a Bloch vector model to describe the quantum state
interference of a single-atom wavefunction in a time-variable
asymmetric double-well potential. The probability of finding the
atom in the first excited state is closely associated with the
magnitude of the spatially dependent external potential and could be
used as a measure of the applied asymmetry. Transitions between
ground and excited states occur during the splitting and
recombination stages. Larmor precession of the Bloch vector during
the holding stage is induced by the asymmetry, will effect an
interferometric phase and determine the final value of the excited
state population. The evolution of the Bloch vector during the
splitting and merging stages is also important because it will
affect the measurable probability $P_{1}$. We have also shown that
special requirements apply to the duration of the splitting and
merging stages in order to avoid excitation of higher modes for
short times and partial adiabatic following if the splitting is too
long. Both these effects lead to a decrease of the measured signal.
Interestingly enough they do not affect the contrast of the
interference fringes if the first excited state is not initially
populated.

Adiabatic evolution of the Bloch vector can offer a new way to
measure the first excited state population after the double-well
interferometric process. We have already mentioned that in the
far-split regime the excited state wavefunction does not overlap
with the ground state wavefunction and will predominantly occupy the
higher energy well [Fig.~\ref{Fig2}(c)]. If at the end of the
non-adiabatic splitting, phase evolution and non-adiabatic
recombination process we also add an additional stage of adiabatic
splitting in a known asymmetrical potential, then the wavefunctions
of the two states will be spatially separated. For recording the
output $P_{1}$ we now simply measure the population of the higher
energy well. To shorten the adiabatic evolution time we need to
apply the highest available asymmetry (Fig.~\ref{Fig6}).

\begin{acknowledgments}
We thank T.D. Kieu for fruitful discussions, and T. Vaughan and P.
Drummond for the introduction to the XMDS code. This work has been
supported by the ARC Centre of Excellence for Quantum-Atom Optics.
\end{acknowledgments}

\appendix*

\section{Derivation of Bloch Equations}

The state vectors at time $t$ and at time $0$ are related via the
unitary evolution operator $\widehat{U(t)}$ as $\vert \Psi(t)\rangle
=\widehat{U(t)}\vert \Psi(0)\rangle$. Operators in the Heisenberg
and Schr\"{o}dinger pictures are related via $\widehat{U}$ as
$(\widehat{\Omega})^{H}=(\widehat{U})^{\dag}(\widehat{\Omega
})^{S}(\widehat{U})$. The expectation values of operators in the two
pictures are related as $\langle \widehat{\Omega}\rangle
=\langle\Psi(t)|(\widehat{\Omega})^{S}|\Psi(t) \rangle =\langle\Psi
(0)|(\widehat{\Omega})^{H}|\Psi (0)\rangle$.

The equation of motion for the Bloch vector components can be
derived using the Heisenberg picture via
\begin{equation}
\frac{d}{dt}\sigma_{a}=\langle \Psi (0)|
\frac{d}{dt}(\widehat{\sigma_a})^{H}|\Psi(0)\rangle,
\end{equation}
where the Heisenberg equation of motion for the Pauli spin operator
in dimensionless units is
\begin{equation}
\frac{d}{dt}\,(\widehat{\sigma_a})^{H}= -i\,[\,(\widehat{\sigma_a})^{H}, (\widehat{H}\mathbf{\,)}^{H}%
\mathbf{\,{]+}}\left(\frac{\partial}{\partial
t}\,(\widehat{\sigma_a})\right)^{H}. \label{Eq.Heisenberg}
\end{equation}

The derivation involves the use of the commutation rules for the
Pauli spin operators. For the first term, we have after substituting
for $\widehat{H}$ from Eq. (\ref{Eq. Ham Opr Pauli})
\begin{eqnarray}
-i\,[(\widehat{\sigma_a})^{H},(\widehat{H})^{H}]
&=&\frac{-i}{2}\left({\Omega_0[\widehat{\sigma_a},\widehat{1}]
+\sum_{{b=}{x,y,z}}\Omega_b [\widehat{\sigma_a},\widehat{\sigma_b}]
}\right)^{H} \nonumber\\
&=& \left(\vec{\Omega}\times(\vec{\widehat{\sigma}})^{H}\right)_{a}.
\end{eqnarray}

Hence the contribution from the first term in Eq.
(\ref{Eq.Heisenberg}) is given by
\begin{equation}
\langle\,\Psi(0)\,|\,\frac{d}{dt}(\widehat{\sigma_a}%
)^{H}\,|\,\Psi (0)\rangle_{1}=\left( \vec{\Omega} \times
\vec{\sigma}\right)_{a}.
\end{equation}

For the contribution from the second term in Eq.
(\ref{Eq.Heisenberg}), we may first write $\widehat{\sigma_a}
=\sum\limits_{A,B=L,R} K_{AB}^{a}\,|A\rangle \,\langle B|$, where
the $K_{AB}^{a}$ are time independent coefficients that can be read
from Eqs. (\ref{Eq. Sxyz}), and then
\begin{eqnarray}
\left(\frac{\partial}{\partial t}(\widehat{\sigma_a})\right)^{H}
&=&\sum\limits_{A,B=L,R} K_{AB}^{a} \nonumber\\
&&\times \left[(\frac{\partial}
 {\partial t}|A\rangle)\langle B| + \vert A\rangle
 (\frac{\partial}{\partial t}\langle B|)
 \right]^{H} .
\end{eqnarray}

Using Eq. (\ref{Eq.State Vector Expn}) for the state vector and
reverting to the Schr\"{o}dinger picture we have
\begin{widetext}
\begin{eqnarray}
\langle \Psi(0)|\left(\partial_{t}(\widehat{\sigma_a})\right)^{H}
|\Psi(0)\rangle_{2}
&=&\langle\Psi(t)|\left\{\sum\limits_{A,B=L,R}K_{AB}^{a}
 \left[ (\partial_{t}|A\rangle )\langle B|+ |A\rangle(\partial_{t}\langle B|)
 \right]\right\}^{S} | \Psi(t)\rangle \nonumber\\
&=& \sum\limits_{A,B,D=L,R}K_{AB}^{a}\left[ C_{D}^{\ast}C_{B}
\langle D|(\partial_{t}|A\rangle) +
C_{A}^{\ast}C_{D}(\partial_{t}\langle B|)|D\rangle \right] .
\end{eqnarray}
\end{widetext}

To evaluate this result, a consideration of the four quantities
$\langle A|\,(\partial_{t}|B\rangle )$, where $(A,\,B=L,\,R)$ is
required, noting also that $(\partial_{t}\langle B|)\,|A\rangle
=(\langle A|\,(\partial_{t}|B\rangle))^{\ast}$. These four
quantities can be expressed in terms of related matrix elements in
the symmetric basis $\langle S_{i}|\,(\partial_{t}|S_{j}\rangle)$,
where $(i,\,j=0,\,1)$. Note $\vert S_{0}\rangle \equiv \vert
S\rangle$ and $\vert S_{1}\rangle \equiv \vert AS\rangle$.

Using the normalisation and reality property, we first show that the
diagonal terms $\langle S_{i}|\,(\partial_{t}|S_{i}\rangle)$ are
zero. For the off-diagonal terms $\langle S_{i}|\,
(\partial_{t}|S_{j}\rangle)$, these are zero because $|S_{i}\rangle$
and $\partial_{t}|S_{j}\rangle$ have opposite symmetry. From these
considerations we
find that all matrix elements $\langle A|\,(\partial_{t%
}|B\rangle)$, where $(A,\,B=L,\,R)$, are zero. Hence the second
contribution to the equation of motion for the Bloch vector
component is zero
\begin{equation}
\langle\,\Psi(0)\,|\,\frac{d}{dt}(\widehat{\sigma_{a}}%
)^{H}\,|\,\Psi (0)\rangle _{2}=0.
\end{equation}

Combining both contributions we find that
\begin{equation}
\frac{d}{dt}\sigma_a=\left( \vec{\Omega} \times \vec{\sigma} \right)
_{a}.
\end{equation}
Thus the Bloch equations can be expressed in vector form as in
Eq.(\ref{Eq. Bloch Eqns Torque}) and in detail as in Eqs. (\ref{Eq.
Bloch Eqn}).

%



\bibliographystyle{apsrev}
\bibliography{DWAI}

\begin{thebibliography}{30}
\expandafter\ifx\csname natexlab\endcsname\relax\def\natexlab#1{#1}\fi
\expandafter\ifx\csname bibnamefont\endcsname\relax
  \def\bibnamefont#1{#1}\fi
\expandafter\ifx\csname bibfnamefont\endcsname\relax
  \def\bibfnamefont#1{#1}\fi
\expandafter\ifx\csname citenamefont\endcsname\relax
  \def\citenamefont#1{#1}\fi
\expandafter\ifx\csname url\endcsname\relax
  \def\url#1{\texttt{#1}}\fi
\expandafter\ifx\csname urlprefix\endcsname\relax\def\urlprefix{URL }\fi
\providecommand{\bibinfo}[2]{#2}
\providecommand{\eprint}[2][]{\url{#2}}

\bibitem[{\citenamefont{Javanainen}(1986)}]{Javanainen86}
\bibinfo{author}{\bibfnamefont{J.}~\bibnamefont{Javanainen}},
  \bibinfo{journal}{Phys. Rev. Lett.} \textbf{\bibinfo{volume}{57}},
  \bibinfo{pages}{3164} (\bibinfo{year}{1986}).

\bibitem[{\citenamefont{Jack et~al.}(1996)\citenamefont{Jack, Collett, and
  Walls}}]{Jack96}
\bibinfo{author}{\bibfnamefont{M.~W.} \bibnamefont{Jack}},
  \bibinfo{author}{\bibfnamefont{M.~J.} \bibnamefont{Collett}},
  \bibnamefont{and} \bibinfo{author}{\bibfnamefont{D.~F.} \bibnamefont{Walls}},
  \bibinfo{journal}{Phys. Rev. A} \textbf{\bibinfo{volume}{54}},
  \bibinfo{pages}{R4625} (\bibinfo{year}{1996}).

\bibitem[{\citenamefont{Menotti et~al.}(2001)\citenamefont{Menotti, Anglin,
  Cirac, and Zoller}}]{Menotti01}
\bibinfo{author}{\bibfnamefont{C.}~\bibnamefont{Menotti}},
  \bibinfo{author}{\bibfnamefont{J.~R.} \bibnamefont{Anglin}},
  \bibinfo{author}{\bibfnamefont{J.~I.} \bibnamefont{Cirac}}, \bibnamefont{and}
  \bibinfo{author}{\bibfnamefont{P.}~\bibnamefont{Zoller}},
  \bibinfo{journal}{Phys. Rev. A} \textbf{\bibinfo{volume}{63}},
  \bibinfo{pages}{023601} (\bibinfo{year}{2001}).

\bibitem[{\citenamefont{Castin and Dalibard}(1997)}]{Castin97}
\bibinfo{author}{\bibfnamefont{Y.}~\bibnamefont{Castin}} \bibnamefont{and}
  \bibinfo{author}{\bibfnamefont{J.}~\bibnamefont{Dalibard}},
  \bibinfo{journal}{Phys. Rev. A} \textbf{\bibinfo{volume}{55}},
  \bibinfo{pages}{4330} (\bibinfo{year}{1997}).

\bibitem[{\citenamefont{Javanainen and Yoo}(1996)}]{Javanainen96}
\bibinfo{author}{\bibfnamefont{J.}~\bibnamefont{Javanainen}} \bibnamefont{and}
  \bibinfo{author}{\bibfnamefont{S.~M.} \bibnamefont{Yoo}},
  \bibinfo{journal}{Phys. Rev. Lett.} \textbf{\bibinfo{volume}{76}},
  \bibinfo{pages}{161} (\bibinfo{year}{1996}).

\bibitem[{\citenamefont{Folman et~al.}(2002)\citenamefont{Folman, Kr{\"{u}}ger,
  Schmiedmayer, Denschlag, and Henkel}}]{Folman02}
\bibinfo{author}{\bibfnamefont{R.}~\bibnamefont{Folman}},
  \bibinfo{author}{\bibfnamefont{P.}~\bibnamefont{Kr{\"{u}}ger}},
  \bibinfo{author}{\bibfnamefont{J.}~\bibnamefont{Schmiedmayer}},
  \bibinfo{author}{\bibfnamefont{J.}~\bibnamefont{Denschlag}},
  \bibnamefont{and} \bibinfo{author}{\bibfnamefont{C.}~\bibnamefont{Henkel}},
  \bibinfo{journal}{Adv. in Atom., Mol. and Opt. Phys.}
  \textbf{\bibinfo{volume}{48}}, \bibinfo{pages}{263} (\bibinfo{year}{2002}).

\bibitem[{\citenamefont{H{\"{a}}nsel
  et~al.}(2001{\natexlab{a}})\citenamefont{H{\"{a}}nsel, Hommelhoff,
  H{\"{a}}nsch, and Reichel}}]{Haensel01a}
\bibinfo{author}{\bibfnamefont{W.}~\bibnamefont{H{\"{a}}nsel}},
  \bibinfo{author}{\bibfnamefont{P.}~\bibnamefont{Hommelhoff}},
  \bibinfo{author}{\bibfnamefont{T.~W.} \bibnamefont{H{\"{a}}nsch}},
  \bibnamefont{and} \bibinfo{author}{\bibfnamefont{J.}~\bibnamefont{Reichel}},
  \bibinfo{journal}{Nature} \textbf{\bibinfo{volume}{413}},
  \bibinfo{pages}{498} (\bibinfo{year}{2001}{\natexlab{a}}).

\bibitem[{\citenamefont{Ott et~al.}(2001)\citenamefont{Ott, Fortagh,
  Schlotterbeck, Grossmann, and Zimmermann}}]{Ott01}
\bibinfo{author}{\bibfnamefont{H.}~\bibnamefont{Ott}},
  \bibinfo{author}{\bibfnamefont{J.}~\bibnamefont{Fortagh}},
  \bibinfo{author}{\bibfnamefont{G.}~\bibnamefont{Schlotterbeck}},
  \bibinfo{author}{\bibfnamefont{A.}~\bibnamefont{Grossmann}},
  \bibnamefont{and}
  \bibinfo{author}{\bibfnamefont{C.}~\bibnamefont{Zimmermann}},
  \bibinfo{journal}{Phys. Rev. Lett.} \textbf{\bibinfo{volume}{87}},
  \bibinfo{pages}{230401} (\bibinfo{year}{2001}).

\bibitem[{\citenamefont{Dumke et~al.}(2002)\citenamefont{Dumke, Volk,
  M{\"{u}}ther, Buchkremer, Birkl, and Ertmer}}]{Dumke02}
\bibinfo{author}{\bibfnamefont{R.}~\bibnamefont{Dumke}},
  \bibinfo{author}{\bibfnamefont{M.}~\bibnamefont{Volk}},
  \bibinfo{author}{\bibfnamefont{T.}~\bibnamefont{M{\"{u}}ther}},
  \bibinfo{author}{\bibfnamefont{F.~B.~J.} \bibnamefont{Buchkremer}},
  \bibinfo{author}{\bibfnamefont{G.}~\bibnamefont{Birkl}}, \bibnamefont{and}
  \bibinfo{author}{\bibfnamefont{W.}~\bibnamefont{Ertmer}},
  \bibinfo{journal}{Phys. Rev. Lett.} \textbf{\bibinfo{volume}{89}},
  \bibinfo{pages}{097903} (\bibinfo{year}{2002}).

\bibitem[{\citenamefont{Hinds et~al.}(2001)\citenamefont{Hinds, Vale, and
  Boshier}}]{Hinds01}
\bibinfo{author}{\bibfnamefont{E.~A.} \bibnamefont{Hinds}},
  \bibinfo{author}{\bibfnamefont{C.~J.} \bibnamefont{Vale}}, \bibnamefont{and}
  \bibinfo{author}{\bibfnamefont{M.~G.} \bibnamefont{Boshier}},
  \bibinfo{journal}{Phys. Rev. Lett.} \textbf{\bibinfo{volume}{86}},
  \bibinfo{pages}{1462} (\bibinfo{year}{2001}).

\bibitem[{\citenamefont{H{\"{a}}nsel
  et~al.}(2001{\natexlab{b}})\citenamefont{H{\"{a}}nsel, Reichel, Hommelhoff,
  and H{\"{a}}nsch}}]{Haensel01b}
\bibinfo{author}{\bibfnamefont{W.}~\bibnamefont{H{\"{a}}nsel}},
  \bibinfo{author}{\bibfnamefont{J.}~\bibnamefont{Reichel}},
  \bibinfo{author}{\bibfnamefont{P.}~\bibnamefont{Hommelhoff}},
  \bibnamefont{and} \bibinfo{author}{\bibfnamefont{T.~W.}
  \bibnamefont{H{\"{a}}nsch}}, \bibinfo{journal}{Phys. Rev. A}
  \textbf{\bibinfo{volume}{64}}, \bibinfo{pages}{063607}
  (\bibinfo{year}{2001}{\natexlab{b}}).

\bibitem[{\citenamefont{Andersson et~al.}(2002)\citenamefont{Andersson,
  Calarco, Folman, Andersson, Hessmo, and Schmiedmayer}}]{Andersson02}
\bibinfo{author}{\bibfnamefont{E.}~\bibnamefont{Andersson}},
  \bibinfo{author}{\bibfnamefont{T.}~\bibnamefont{Calarco}},
  \bibinfo{author}{\bibfnamefont{R.}~\bibnamefont{Folman}},
  \bibinfo{author}{\bibfnamefont{M.}~\bibnamefont{Andersson}},
  \bibinfo{author}{\bibfnamefont{B.}~\bibnamefont{Hessmo}}, \bibnamefont{and}
  \bibinfo{author}{\bibfnamefont{J.}~\bibnamefont{Schmiedmayer}},
  \bibinfo{journal}{Phys. Rev. Lett.} \textbf{\bibinfo{volume}{88}},
  \bibinfo{pages}{100401} (\bibinfo{year}{2002}).

\bibitem[{\citenamefont{Stickney and Zozulya}(2002)}]{Stickney02}
\bibinfo{author}{\bibfnamefont{J.~A.} \bibnamefont{Stickney}} \bibnamefont{and}
  \bibinfo{author}{\bibfnamefont{A.~A.} \bibnamefont{Zozulya}},
  \bibinfo{journal}{Phys. Rev. A} \textbf{\bibinfo{volume}{66}},
  \bibinfo{pages}{053601} (\bibinfo{year}{2002}).

\bibitem[{\citenamefont{Negretti and Henkel}(2004)}]{Negretti04}
\bibinfo{author}{\bibfnamefont{A.}~\bibnamefont{Negretti}} \bibnamefont{and}
  \bibinfo{author}{\bibfnamefont{C.}~\bibnamefont{Henkel}},
  \bibinfo{journal}{J. Phys. B} \textbf{\bibinfo{volume}{37}},
  \bibinfo{pages}{385} (\bibinfo{year}{2004}).

\bibitem[{\citenamefont{Kreutzmann et~al.}(2004)\citenamefont{Kreutzmann,
  Poulsen, Lewenstein, Dumke, Ertmer, Birkl, and Sanpera}}]{Kreutzmann04}
\bibinfo{author}{\bibfnamefont{H.}~\bibnamefont{Kreutzmann}},
  \bibinfo{author}{\bibfnamefont{U.~V.} \bibnamefont{Poulsen}},
  \bibinfo{author}{\bibfnamefont{M.}~\bibnamefont{Lewenstein}},
  \bibinfo{author}{\bibfnamefont{R.}~\bibnamefont{Dumke}},
  \bibinfo{author}{\bibfnamefont{W.}~\bibnamefont{Ertmer}},
  \bibinfo{author}{\bibfnamefont{G.}~\bibnamefont{Birkl}}, \bibnamefont{and}
  \bibinfo{author}{\bibfnamefont{A.}~\bibnamefont{Sanpera}},
  \bibinfo{journal}{Phys. Rev. Lett.} \textbf{\bibinfo{volume}{92}},
  \bibinfo{pages}{163201} (\bibinfo{year}{2004}).

\bibitem[{\citenamefont{Cassettari et~al.}(2000)\citenamefont{Cassettari,
  Hessmo, Folman, Maier, and Schmiedmayer}}]{Cassettari00}
\bibinfo{author}{\bibfnamefont{D.}~\bibnamefont{Cassettari}},
  \bibinfo{author}{\bibfnamefont{B.}~\bibnamefont{Hessmo}},
  \bibinfo{author}{\bibfnamefont{R.}~\bibnamefont{Folman}},
  \bibinfo{author}{\bibfnamefont{T.}~\bibnamefont{Maier}}, \bibnamefont{and}
  \bibinfo{author}{\bibfnamefont{J.}~\bibnamefont{Schmiedmayer}},
  \bibinfo{journal}{Phys. Rev. Lett.} \textbf{\bibinfo{volume}{85}},
  \bibinfo{pages}{5483} (\bibinfo{year}{2000}).

\bibitem[{\citenamefont{M{\"{u}}ller et~al.}(2000)\citenamefont{M{\"{u}}ller,
  Cornell, Prevedelli, Schwindt, Zozulya, and Anderson}}]{Muller00}
\bibinfo{author}{\bibfnamefont{D.}~\bibnamefont{M{\"{u}}ller}},
  \bibinfo{author}{\bibfnamefont{E.~A.} \bibnamefont{Cornell}},
  \bibinfo{author}{\bibfnamefont{M.}~\bibnamefont{Prevedelli}},
  \bibinfo{author}{\bibfnamefont{P.~D.~D.} \bibnamefont{Schwindt}},
  \bibinfo{author}{\bibfnamefont{A.}~\bibnamefont{Zozulya}}, \bibnamefont{and}
  \bibinfo{author}{\bibfnamefont{D.~Z.} \bibnamefont{Anderson}},
  \bibinfo{journal}{Opt. Lett.} \textbf{\bibinfo{volume}{25}},
  \bibinfo{pages}{1382} (\bibinfo{year}{2000}).

\bibitem[{\citenamefont{Est{\`{e}}ve et~al.}(2005)\citenamefont{Est{\`{e}}ve,
  Schumm, Trebbia, Bouchoule, Aspect, and Westbrook}}]{Esteve05}
\bibinfo{author}{\bibfnamefont{J.}~\bibnamefont{Est{\`{e}}ve}},
  \bibinfo{author}{\bibfnamefont{T.}~\bibnamefont{Schumm}},
  \bibinfo{author}{\bibfnamefont{J.-B.} \bibnamefont{Trebbia}},
  \bibinfo{author}{\bibfnamefont{I.}~\bibnamefont{Bouchoule}},
  \bibinfo{author}{\bibfnamefont{A.}~\bibnamefont{Aspect}}, \bibnamefont{and}
  \bibinfo{author}{\bibfnamefont{C.~I.} \bibnamefont{Westbrook}},
  \bibinfo{journal}{Eur. Phys. J. D} \textbf{\bibinfo{volume}{35}},
  \bibinfo{pages}{141} (\bibinfo{year}{2005}).

\bibitem[{\citenamefont{H{\"{a}}nsel
  et~al.}(2001{\natexlab{c}})\citenamefont{H{\"{a}}nsel, Reichel, Hommelhoff,
  and H{\"{a}}nsch}}]{Haensel01c}
\bibinfo{author}{\bibfnamefont{W.}~\bibnamefont{H{\"{a}}nsel}},
  \bibinfo{author}{\bibfnamefont{J.}~\bibnamefont{Reichel}},
  \bibinfo{author}{\bibfnamefont{P.}~\bibnamefont{Hommelhoff}},
  \bibnamefont{and} \bibinfo{author}{\bibfnamefont{T.~W.}
  \bibnamefont{H{\"{a}}nsch}}, \bibinfo{journal}{Phys. Rev. Lett.}
  \textbf{\bibinfo{volume}{86}}, \bibinfo{pages}{608}
  (\bibinfo{year}{2001}{\natexlab{c}}).

\bibitem[{\citenamefont{Shin et~al.}(2004)\citenamefont{Shin, Saba, Pasquini,
  Ketterle, Pritchard, and Leanhardt}}]{Shin04}
\bibinfo{author}{\bibfnamefont{Y.}~\bibnamefont{Shin}},
  \bibinfo{author}{\bibfnamefont{M.}~\bibnamefont{Saba}},
  \bibinfo{author}{\bibfnamefont{T.~A.} \bibnamefont{Pasquini}},
  \bibinfo{author}{\bibfnamefont{W.}~\bibnamefont{Ketterle}},
  \bibinfo{author}{\bibfnamefont{D.~E.} \bibnamefont{Pritchard}},
  \bibnamefont{and} \bibinfo{author}{\bibfnamefont{A.~E.}
  \bibnamefont{Leanhardt}}, \bibinfo{journal}{Phys. Rev. Lett.}
  \textbf{\bibinfo{volume}{92}}, \bibinfo{pages}{050405}
  (\bibinfo{year}{2004}).

\bibitem[{\citenamefont{Schumm et~al.}(2005)\citenamefont{Schumm, Hofferberth,
  Andersson, Wildermuth, Groth, Bar-Joseph, Schmiedmayer, and
  Kr{\"{u}}ger}}]{Schumm05}
\bibinfo{author}{\bibfnamefont{T.}~\bibnamefont{Schumm}},
  \bibinfo{author}{\bibfnamefont{S.}~\bibnamefont{Hofferberth}},
  \bibinfo{author}{\bibfnamefont{L.~M.} \bibnamefont{Andersson}},
  \bibinfo{author}{\bibfnamefont{S.}~\bibnamefont{Wildermuth}},
  \bibinfo{author}{\bibfnamefont{S.}~\bibnamefont{Groth}},
  \bibinfo{author}{\bibfnamefont{I.}~\bibnamefont{Bar-Joseph}},
  \bibinfo{author}{\bibfnamefont{J.}~\bibnamefont{Schmiedmayer}},
  \bibnamefont{and}
  \bibinfo{author}{\bibfnamefont{P.}~\bibnamefont{Kr{\"{u}}ger}},
  \bibinfo{journal}{Nature Physics} \textbf{\bibinfo{volume}{1}},
  \bibinfo{pages}{57} (\bibinfo{year}{2005}).

\bibitem[{\citenamefont{Shin et~al.}(2005)\citenamefont{Shin, Sanner, Jo,
  Pasquini, Saba, Ketterle, Pritchard, Vengalattore, and Prentiss}}]{Shin05}
\bibinfo{author}{\bibfnamefont{Y.}~\bibnamefont{Shin}},
  \bibinfo{author}{\bibfnamefont{C.}~\bibnamefont{Sanner}},
  \bibinfo{author}{\bibfnamefont{G.-B.} \bibnamefont{Jo}},
  \bibinfo{author}{\bibfnamefont{T.~A.} \bibnamefont{Pasquini}},
  \bibinfo{author}{\bibfnamefont{M.}~\bibnamefont{Saba}},
  \bibinfo{author}{\bibfnamefont{W.}~\bibnamefont{Ketterle}},
  \bibinfo{author}{\bibfnamefont{D.~E.} \bibnamefont{Pritchard}},
  \bibinfo{author}{\bibfnamefont{M.}~\bibnamefont{Vengalattore}},
  \bibnamefont{and} \bibinfo{author}{\bibfnamefont{M.}~\bibnamefont{Prentiss}},
  \bibinfo{journal}{Phys. Rev. A} \textbf{\bibinfo{volume}{72}},
  \bibinfo{pages}{021604(R)} (\bibinfo{year}{2005}).

\bibitem[{\citenamefont{Kasevich}(2002)}]{Kasevich02}
\bibinfo{author}{\bibfnamefont{M.~A.} \bibnamefont{Kasevich}},
  \bibinfo{journal}{Science} \textbf{\bibinfo{volume}{298}},
  \bibinfo{pages}{1363} (\bibinfo{year}{2002}).

\bibitem[{\citenamefont{Javanainen and Wilkens}(1997)}]{Javanainen97}
\bibinfo{author}{\bibfnamefont{J.}~\bibnamefont{Javanainen}} \bibnamefont{and}
  \bibinfo{author}{\bibfnamefont{M.}~\bibnamefont{Wilkens}},
  \bibinfo{journal}{Phys. Rev. Lett.} \textbf{\bibinfo{volume}{78}},
  \bibinfo{pages}{4675} (\bibinfo{year}{1997}).

\bibitem[{\citenamefont{Calarco et~al.}(2000)\citenamefont{Calarco, Hinds,
  Jaksch, Schmiedmayer, Cirac, and Zoller}}]{Calarco00}
\bibinfo{author}{\bibfnamefont{T.}~\bibnamefont{Calarco}},
  \bibinfo{author}{\bibfnamefont{E.~A.} \bibnamefont{Hinds}},
  \bibinfo{author}{\bibfnamefont{D.}~\bibnamefont{Jaksch}},
  \bibinfo{author}{\bibfnamefont{J.}~\bibnamefont{Schmiedmayer}},
  \bibinfo{author}{\bibfnamefont{J.~I.} \bibnamefont{Cirac}}, \bibnamefont{and}
  \bibinfo{author}{\bibfnamefont{P.}~\bibnamefont{Zoller}},
  \bibinfo{journal}{Phys. Rev. A} \textbf{\bibinfo{volume}{61}},
  \bibinfo{pages}{022304} (\bibinfo{year}{2000}).

\bibitem[{\citenamefont{Pulido}(2003)}]{Pulido03}
\bibinfo{author}{\bibfnamefont{D.}~\bibnamefont{Pulido}},
  \emph{\bibinfo{title}{Instability in a cold atom interferometer}},
  \bibinfo{howpublished}{Master thesis, Worcester Politechnic Institute}
  (\bibinfo{year}{2003}).

\bibitem[{\citenamefont{Berry}(1984)}]{Berry84}
\bibinfo{author}{\bibfnamefont{M.~V.} \bibnamefont{Berry}},
  \bibinfo{journal}{Proc. Roy. Soc. London, Ser. A}
  \textbf{\bibinfo{volume}{392}}, \bibinfo{pages}{45} (\bibinfo{year}{1984}).

\bibitem[{\citenamefont{Allen and Eberly}(1975)}]{Allen75}
\bibinfo{author}{\bibfnamefont{L.}~\bibnamefont{Allen}} \bibnamefont{and}
  \bibinfo{author}{\bibfnamefont{J.~H.} \bibnamefont{Eberly}},
  \emph{\bibinfo{title}{Optical Resonance and Two-Level Atoms}}
  (\bibinfo{publisher}{New York: Wiley}, \bibinfo{year}{1975}).

\bibitem[{\citenamefont{Cohen-Tannoudji
  et~al.}(1977)\citenamefont{Cohen-Tannoudji, Diu, and
  Lalo{\"{e}}}}]{Tannoudji77}
\bibinfo{author}{\bibfnamefont{C.}~\bibnamefont{Cohen-Tannoudji}},
  \bibinfo{author}{\bibfnamefont{B.}~\bibnamefont{Diu}}, \bibnamefont{and}
  \bibinfo{author}{\bibfnamefont{F.}~\bibnamefont{Lalo{\"{e}}}},
  \emph{\bibinfo{title}{Quantum Mechanics, Volume I}} (\bibinfo{publisher}{New
  York: Wiley}, \bibinfo{year}{1977}).

\bibitem[{XMD()}]{XMDS}
\bibinfo{howpublished}{XMDS code is a code generator that integrates equations.
  It is developed at the University of Queensland, Brisbane, Australia},
  \eprint{www.xmds.org}.

\end{thebibliography}

\end{document}